\documentclass[aps,prl,twocolumn,superscriptaddress,reprint,nobibnotes]{revtex4-2}

\makeatletter
\def\NAT@cmprs{\z@} 
\makeatother

\usepackage{amsmath}
\usepackage{amsfonts}
\usepackage{graphicx}
\usepackage{cancel}

\usepackage[colorinlistoftodos,prependcaption,textsize=small]{todonotes}
\usepackage{color}
\definecolor{red}{rgb}{0.75,0,0}
\definecolor{blue}{rgb}{0,0,0.75}
\definecolor{green}{rgb}{0,0.5,0}
\definecolor{orange-red}{RGB}{255,69,0}

\makeatletter

\newcommand{\eq}{\begin{equation}}
\newcommand{\eeq}{\end{equation}}
\newcommand{\aeq}{\begin{equation}\begin{aligned}}
\newcommand{\eaeq}{\end{aligned}\end{equation}}

\makeatother

\begin{document}

\title{Structured flows regulate droplet dynamics in phase-separating fluids}

\author{Yulin Li}
\thanks{These authors contributed equally}
\affiliation{Fujian Provincial Key Laboratory for Soft Functional Materials Research, Research Institute for Biomimetics and Soft Matter, Department of Physics, Xiamen University, Xiamen, Fujian 361005, China}

\author{Tong Zhou}
\thanks{These authors contributed equally}
\affiliation{Fujian Provincial Key Laboratory for Soft Functional Materials Research, Research Institute for Biomimetics and Soft Matter, Department of Physics, Xiamen University, Xiamen, Fujian 361005, China}
 
\author{Yanyu Li}
\thanks{These authors contributed equally}
\affiliation{Fujian Provincial Key Laboratory for Soft Functional Materials Research, Research Institute for Biomimetics and Soft Matter, Department of Physics, Xiamen University, Xiamen, Fujian 361005, China}

\author{Qi Zhang}
\affiliation{School of Science, Jiangsu University of Science and Technology, Zhenjiang, Jiangsu 212100, China}

\author{Zhihong You}
\thanks{Corresponding author: zhyou@xmu.edu.cn}
\affiliation{Fujian Provincial Key Laboratory for Soft Functional Materials Research, Research Institute for Biomimetics and Soft Matter, Department of Physics, Xiamen University, Xiamen, Fujian 361005, China}
	
\date{\today}
	
\begin{abstract}
Liquid–liquid phase separation is a fundamental route by which homogeneous fluids self-organize into mesoscale droplets and domains. In passive mixtures, the fate of these structures is largely dictated by curvature-dependent chemical potentials, which drive interfacial relaxation and Ostwald ripening toward progressively coarser morphologies. Here we show that structured flows can drive phase-separating fluids into flow-selected nonequilibrium states, thereby qualitatively rewriting the droplet dynamics. Using an advective Cahn–Hilliard model, we demonstrate that patterned velocity fields deform droplet interfaces along the streamlines of the imposed flow. The resulting curvature variations reshape the chemical-potential landscape that steers diffusive material transport. The resulting fluxes allow structured vortices to capture droplets, select stable droplet sizes, and invert the conventional ripening hierarchy by rendering larger droplets effectively less stable than smaller ones. In many-droplet systems, periodic and disordered vortex fields arrest coarsening and select steady morphologies whose length scales and symmetries are set by the underlying flow structure. Together, these results establish structured flow as a nonequilibrium route for controlling the thermodynamic driving forces that govern phase-separating fluids.
\end{abstract}

\maketitle


\section{Introduction}
Liquid--liquid phase separation (LLPS) is a fundamental pathway by which multicomponent fluids organize into mesoscale droplets, domains, and condensates. It underlies phenomena ranging from membraneless organelles in living cells~\cite{alberti2019considerations,alberti2025current} to adaptive soft materials~\cite{Wang2021liquid,xu2023liquid,fu2024supramolecular}. At the physical level, LLPS converts local molecular interactions and thermodynamic instabilities into structure on much larger length scales~\cite{cahn1958free,bray2002theory,tanaka2012viscoelastic,tanaka2022viscoelastic}. Once droplets or domains form, however, their subsequent evolution in passive mixtures is governed by a comparatively simple thermodynamic bias: interfacial curvature sets the chemical potential, so highly curved domains tend to lose material while less curved ones grow. This bias drives interfacial relaxation, Ostwald ripening, and ultimately progressive coarsening toward macroscopic phase separation~\cite{lifshitz1961kinetics,bray2002theory}. A fundamental problem is therefore whether external driving can move LLPS beyond this passive selection rule, not merely changing how fast phase separation occurs, but altering the thermodynamic biases that determine which droplets grow, shrink, or persist.

External fields have been shown to remodel the thermodynamic landscape of phase-separating mixtures. Patterned light can locally tune interactions or solubility to nucleate, dissolve, or reposition condensates ~\cite{schneider2021liquid,huang2025liquid}, while electric and magnetic fields exploit dielectric or magnetic contrast to bias composition, interfacial morphology, and domain organization~\cite{tsori2007phase,marcus2009phase}. Flow provides a distinct mode of external driving. Rather than imposing a static energetic bias, flow continuously advects material, allowing hydrodynamic transport to compete with diffusion and interfacial relaxation. The role of flow in LLPS has been investigated for decades, beginning largely with steady shear. In such simple flows, phase-separated domains can be stretched, aligned, fragmented, or arrested, leading to anisotropic coarsening, modified growth laws, and steady droplet-size distributions governed by the balance between deformation, breakup, coalescence, and capillary relaxation~\cite{baumberger1991shear,onuki1997phase,berthier2001phase,han2006effect,olmsted2008perspectives,coronas2024stability,suzuki2024arresting}. More recently, active turbulent flows have revealed a more dynamic mode of flow-driven regulation: chaotic stirring continually fragments and redistributes condensates, suppressing macroscopic coarsening and sustaining nonequilibrium steady states with persistent droplet turnover~\cite{singh2019hydrodynamically,bentley2019physical,adkins2022dynamics,caballero2022activity,tayar2023controlling,bhattacharyya2023phase,naz2024self,gulati2025active,laprade2025droplet,berti2005turbulence}. 
These advances establish flow as a powerful nonequilibrium driving mechanism for LLPS, capable of altering coarsening pathways, arresting domain growth, and sustaining droplet turnover. Furthermore, they point to a missing mode of hydrodynamic control: structured flows that lie between simple shear and disordered turbulence. Whether such flows can regulate nonequilibrium phase-separation dynamics with selectivity and control remains largely unexplored.

Here, we address this problem by studying LLPS under prescribed incompressible flows with coherent spatial structure. We demonstrate that structured flows can sculpt the phase field through the geometry and topology of their streamlines. By guiding droplet interfaces along these streamlines, the flow redistributes interfacial curvature and thereby reorganizes the chemical-potential landscape across the mixture. The resulting chemical-potential gradients drive directed diffusive fluxes, allowing structured flows to regulate nonequilibrium phase-separation dynamics with selectivity and control. This mechanism gives rise to a hierarchy of flow-regulated behaviors. For isolated droplets, structured vortices act as traps that capture and pin the phase-separated domain. For interacting droplets, the geometry of the vortices sets the effective chemical potential of each droplet, enabling stable size selection, tunable size ratios, and even a flow-induced flux reversal. In extended systems, periodic and disordered vortex fields suppress macroscopic coarsening and stabilize steady morphologies whose characteristic length scales and symmetries are encoded by the imposed flow. The same principle also modifies the kinetics and degree of phase separation. These results establish structured flow as a hydrodynamic route for selectively regulating LLPS at the level of thermodynamic fluxes, providing a controlled nonequilibrium counterpart to both simple shear and turbulent mixing.

\section{Model}

We consider a minimal two-dimensional model of Cahn-Hilliard phase field $\phi$ advected by a prescribed incompressible flow field $\textbf{v}$ \cite{cahn1958free}:
\begin{equation}
\label{eq:dtphi}
\frac{\partial \phi}{\partial t}+\mathbf{v}\cdot\nabla\phi=M\nabla^2\mu .
\end{equation}
Here $M$ is the mobility and $\mu=\delta F/\delta\phi$ is the chemical potential associated with the Ginzburg--Landau free energy
\begin{equation}
\label{eq:FreeEner}
F=\int_{\mathbf{r}}\left[\frac{a}{4}\phi^2(1-\phi)^2+\frac{k}{2}|\nabla\phi|^2\right]d\mathbf{r}.
\end{equation}
The double-well term favors coexistence of the two phases, $\phi=0$ and $\phi=1$, while the gradient term penalizes interfaces and sets the interfacial tension. We nondimensionalize length by the interface width $\ell=\sqrt{k/a}$, time by the diffusive relaxation time $\tau=k/(Ma^2)$, and energy by the two-dimensional interfacial energy scale $\epsilon=\sqrt{ak}$. Equation~\eqref{eq:dtphi} is solved by finite differences in a square periodic domain. In general, the phase field can feed back on the flow through capillary stresses. Here we neglect this back-coupling in order to isolate the physics of externally imposed structured flows. The validity of this one-way-coupling approximation is rationalized in Sec. S2 of the SI using full Cahn--Hilliard--Navier--Stokes (CHNS) simulations. 

For each simulation, we prescribe a target velocity field $\mathbf{v}^{\mathrm{tar}}(\mathbf{r})=v_0\hat{\mathbf{v}}^{\mathrm{tar}}(\mathbf{r})$, where $v_0$ sets the flow strength and $\hat{\mathbf{v}}^{\mathrm{tar}}(\mathbf{r})$ specifies the spatial profile, normalized such that $\max_{\mathbf{r}}|\hat{\mathbf{v}}^{\mathrm{tar}}(\mathbf{r})|=1$. The velocity field $\mathbf{v}$ is reconstructed from a streamfunction with periodic boundary conditions and is therefore periodic and incompressible. 
For simulations with prescribed droplets, the phase-field profile of a single diffuse circular droplet is defined as $\phi_c(\mathbf{r}; \mathbf{r}_c, R_0)=[1+\tanh(R_0-|\mathbf{r}-\mathbf{r}_c|)]/2$, where $R_0$ and $\mathbf{r}_c$ denote the initial droplet radius and center, respectively. For a two-droplet configuration, the initial condition is $\phi_c(\mathbf{r}; \mathbf{r}_{c,1}, R_{0,1})+\phi_c(\mathbf{r}; \mathbf{r}_{c,2}, R_{0,2})$.
To simulate phase separation, the system is instead initialized from a nearly homogeneous mixture, $\phi(\mathbf{r},0)=\bar{\phi}+\delta\phi(\mathbf{r}),$ where $\bar{\phi}$ is the conserved mean composition and $\delta\phi(\mathbf{r})$ is a small random perturbation. Figure-specific parameters are given in the captions, with further numerical details provided in Secs.~S1 and S7 of the SI.
Taking $U$ and $L_c$ as the characteristic flow speed and droplet size, the relevant dimensionless groups are \[Pe=\frac{UL_c}{Ma},\qquad Cn=\frac{\ell}{L_c}.\] For typical values $U\simeq1$--$3$ and $L_c\simeq16$--$32$, we obtain $Pe\simeq16$--$96$ and $Cn\simeq0.031$--$0.063$, evaluated using the nondimensional quantities in our simulations.

\section{Single-droplet capture}

\begin{figure*}[t]
    \centering
    \includegraphics[width=1.8\columnwidth]{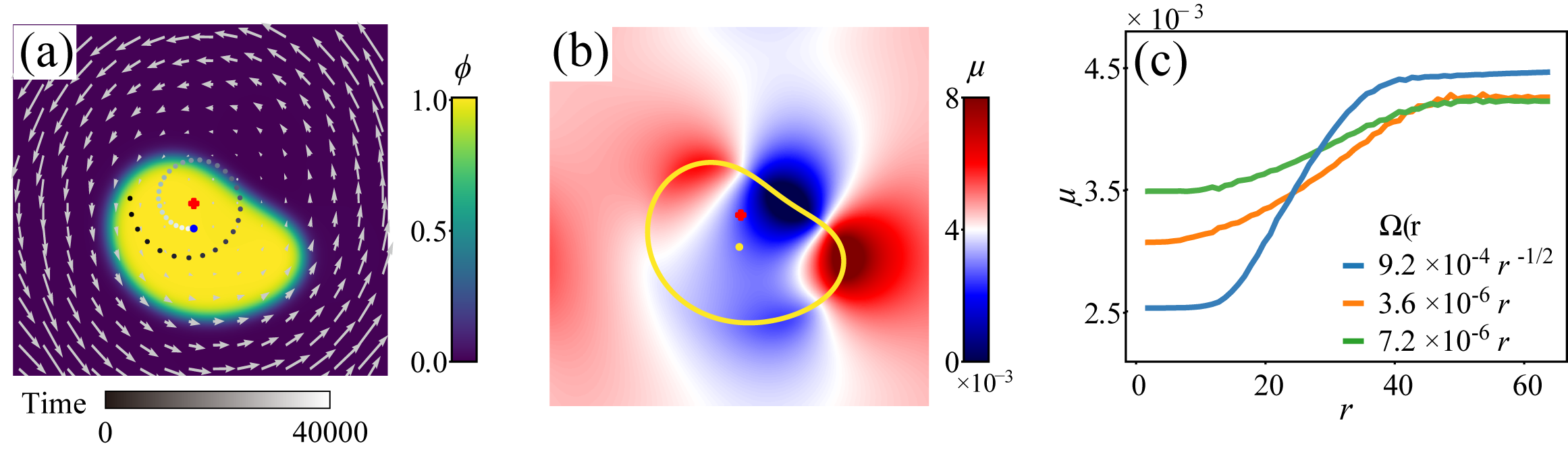} 
    \caption{A droplet captured by a circular vortex with non-uniform angular velocity. (a) Snapshot of the phase field $\phi$ with the velocity field (gray arrows) superimposed on it. The arrow size is proportional to the local speed $|\textbf{v}|$. The dots represent trajectory of the center of mass of the droplet. Here, $\Omega(r)=7.2\times 10^{-6}r$. (b) The spatial distribution of chemical potential corresponding to the phase field in panel a. The orange curve shows the droplet interface. (c) Radial distribution of chemical potential $\langle\mu\rangle$, calculated by averaging $\mu(\textbf{r}')$ within a thin ring $r\le|\textbf{r}'|<r+dr$. Different lines correspond to different $\Omega(r)$ functions. See Sec. S7 for detailed vortex structure.}
    \label{fig:1drop}
\end{figure*}

As a first demonstration, we place a single droplet in a flow that is approximately axisymmetric in the region explored by the droplet. In this region, the angular velocity $\Omega(r)$ depends only on the distance $r$ from the vortex center. Because a perfectly circular vortex is generally incompatible with periodic boundary conditions, the vortices used in our simulations are only approximately circular. When $\Omega$ is uniform, the flow simply rotates the droplet as a rigid body without inducing radial migration (Vid.~1a). We therefore introduce a radial variation in $\Omega(r)$ to generate shear and examine whether flow-induced interfacial deformation can drive droplet migration. In contrast, a nonuniform angular velocity, for example an increasing angular velocity $\Omega(r)\sim r$, causes the droplet center of mass to follow an inward spiral trajectory, until the droplet becomes trapped near the vortex center (Fig. 1a and Vid. 1d). This migration occurs for both signs of the radial variation in angular velocity (Vid. 1b), indicating that it is not caused by direct radial advection, since the imposed velocity field has no radial component. Instead, the effect originates from a flow-induced thermodynamic bias. The nonuniform rotation deforms the droplet interface, redistributing interfacial curvature and thereby reshaping the chemical-potential field (Fig.~\ref{fig:1drop}b). For both types of nonuniform vortex, the interface develops, on average, larger curvature at larger $r$, corresponding to a higher local chemical potential~\cite{sieradzki1993curvature,van1998mean}. The resulting chemical-potential gradient drives a diffusive flux toward the vortex center, transporting droplet material inward (Fig.~\ref{fig:1drop}c). Increasing the radial variation of $\Omega(r)$ strengthens this chemical-potential gradient and accelerates droplet migration (Fig.~\ref{fig:1drop}c and Vid. 1c, 1d).


\begin{figure*}[t]
    \centering
    \includegraphics[width=1.8\columnwidth]{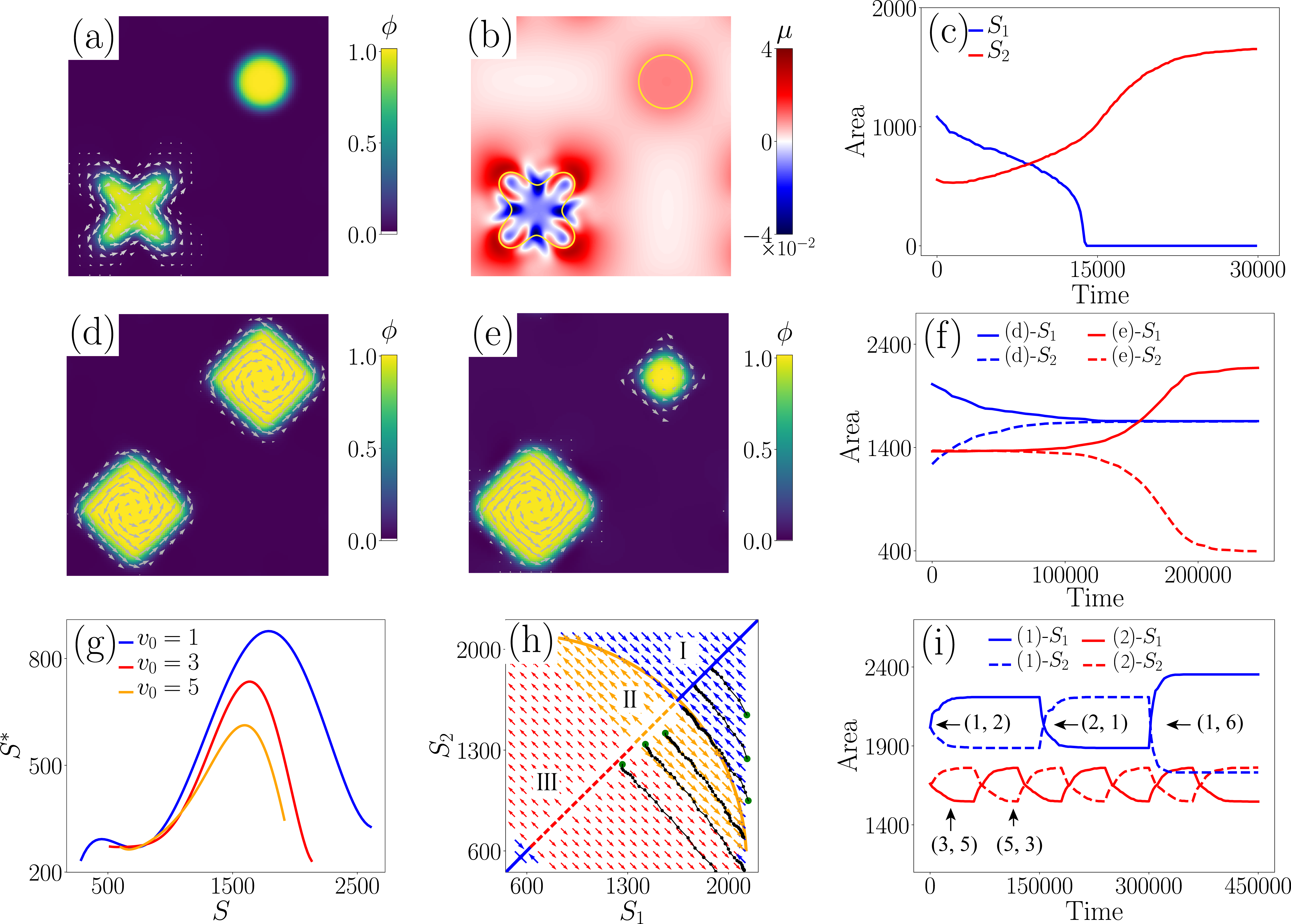}
    \caption{Manipulating the sizes of two droplets using patterned flows. (a-c): A clover-shaped vortex induces a flow-driven reversal of the inter-droplet diffusive flux. Panels a and b show the phase field $\phi$ and the chemical potential $\mu$, respectively. (c) Time evolution of the droplet sizes. (d-e): Stable coexistence of two droplets of (d) equal and (e) different sizes at the steady state. (f) Evolution of the sizes of droplets shown in panels d (blue lines) and e (red lines). (g) Constitutive relations of droplets in square vortices with different $v_0$ values. (h) Phase portrait of two droplets placed in identical square vortices, reconstructed using the \(v_0=3\) constitutive relation (red curve) in panel (g). The arrows indicate the instantaneous direction of evolution predicted by the red constitutive relation in panel g. The solid and dashed lines show stable and unstable fixed points, respectively. The dotted lines show real phase trajectories from a few simulations, starting from the green dots. The small drift of the trajectories is due to the threshold-based extraction of droplet areas, not a violation of mass conservation. (i) The relative sizes of two droplets can be tuned dynamically by changing the flow-strength parameter $v_0$. The ($\cdot$,$\cdot$) indicates the flow-strength parameter $v_0$ of the two square vortices $(v_{01}, v_{02})$. Detailed vortex structures can be found in Sec. S7.B.}
    \label{fig:2drop}
\end{figure*}

\section{Two-droplet dynamics}

The same principle also enables control over the relative sizes of two droplets. In a passive mixture, two-droplet coexistence is intrinsically unstable: the smaller droplet has a higher chemical potential, so diffusive transport transfers material to the larger droplet~\cite{voorhees1992ostwald}. The smaller droplet therefore shrinks and is eventually absorbed, in the familiar process of Ostwald ripening (Vid. 2a)~\cite{voorhees1985theory,taylor1998ostwald}. Structured vortices can reverse this thermodynamic bias. When the larger droplet is placed in a sufficiently strong clover-shaped vortex, the flow pins it near the vortex center and aligns its interface with the streamlines, allowing the imposed geometry to reshape its curvature distribution (Fig.~2a, Fig. S3 and Sec. S4). The enhanced curvature at the petal tips increases the droplet’s mean chemical potential (Fig.~2b), reversing the diffusive flux so that material is transferred to the smaller droplet instead (Fig.~2c and Vid. 2b). More generally, placing both droplets in structured vortices enables finer control of droplet coexistence. Identical vortices tend to equalize droplet sizes (Figs.~2d and 2f, Vid. 2c), whereas vortices with different geometries or strengths can stabilize unequal but well-defined steady sizes (Figs.~2e and 2f, Vid. 2d). These steady sizes can in turn be tuned dynamically by varying the vortex structure or speed.

\begin{figure*}[t]
    \centering
    \includegraphics[width=1.8\columnwidth]{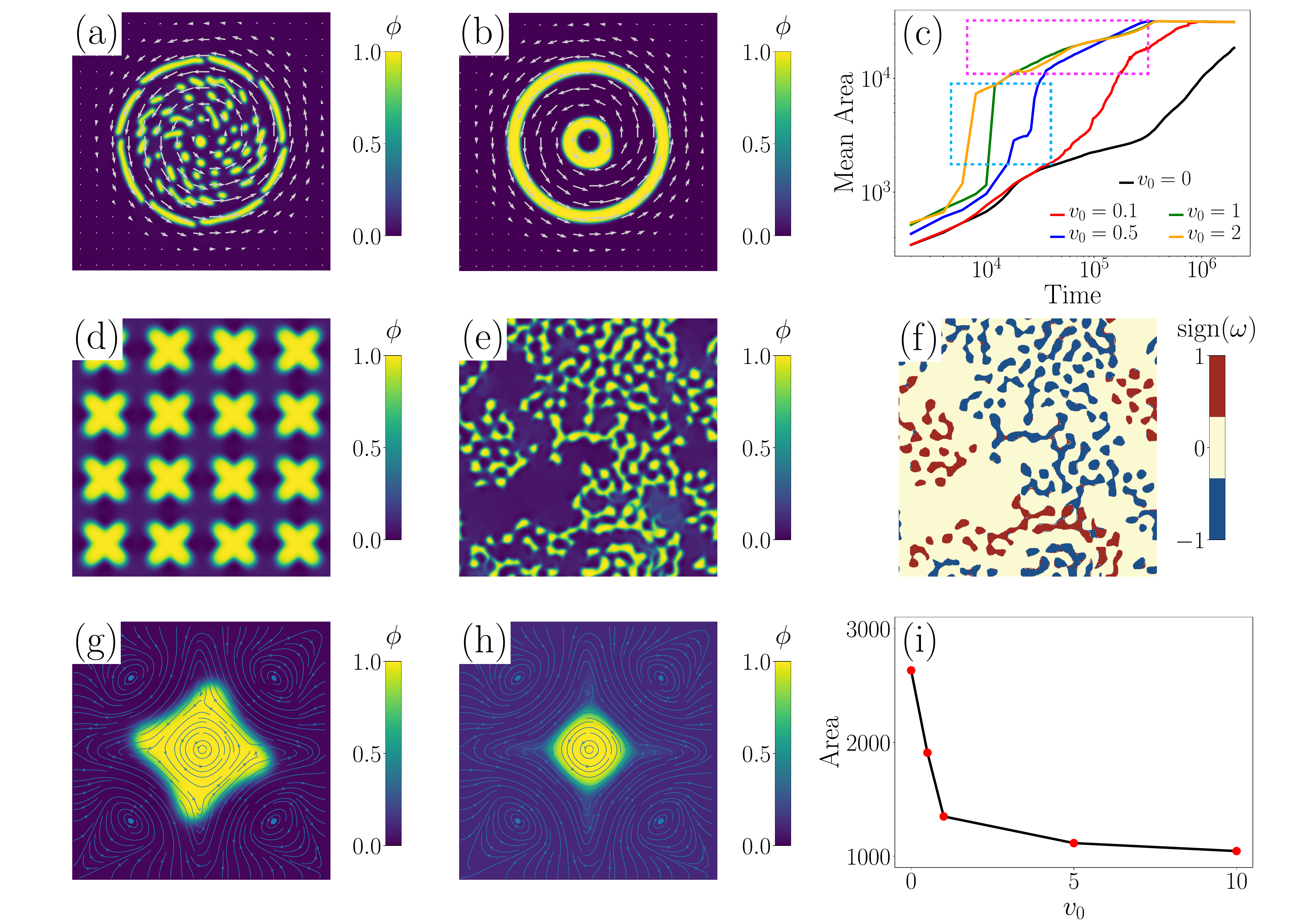}    
    \caption{Patterned flow controls the kinetics, structure, and degree of phase separation. 
(a--b) Phase separation in a circular vortex at (a) $t=5600$ and (b) $t=90800$. 
(c) The area-weighted mean droplet area as a function of time, as defined in Sec.~S3. Different lines correspond to vortices with different flow-strength parameters $v_0$. The cyan and magenta boxes highlight the accelerated and decelerated coarsening regimes. 
(d--e) Steady states of phase separation driven by (d) a periodic array of identical vortices and (e) a steady flow with quenched disorder. 
(f) Same as panel (e), but the droplet color indicates the sign of the local vorticity. 
Panels (a--f) are initialized from $\phi(\mathbf r,0)=\bar{\phi}+\delta\phi(\mathbf r)$, with $\bar{\phi}=0.4$ in (a--d), and $0.3$ in (e--f), where $\delta\phi$ is a small random perturbation. The disordered flow in (e--f) is generated by Fourier filtering a standard-normal random vorticity field, with $k_0=16$ and $\omega_0=0.4$.
(g--h) Steady states initialized from identical circular droplets of radius $R_0=32$ centered at the vortex center, driven by vortices with (g) $v_0=0.01$ and (h) $v_0=10$. 
(i) Steady-state droplet area versus the flow-strength parameter $v_0$. Detailed flow structures are provided in Sec.~S7. C.}
    \label{fig:llps}
\end{figure*}

To rationalize these observations and predict droplet evolution, we introduce an effective chemical potential (ECP) for droplets in structured vortices. For a given vortex, we place a droplet of area $S$ inside the flow and a circular reference droplet in a flow-free region (see, e.g., Fig.~2a). By varying the reference-droplet area, we identify a value $S^*$ at which the net diffusive flux between the two droplets vanishes. We define the effective chemical potential (ECP) of the advected droplet as the equilibrium chemical potential of a passive circular reference droplet of area $S^*$ for which the net diffusive flux between the two droplets vanishes, i.e., $\mu_{\mathrm{eff}}(S, \mathbf{v}) \equiv \mu_{\mathrm{eq}}^{\mathrm{circ}}(S^*)$. Unlike the local equilibrium chemical potential $\mu(\mathbf{r})=\delta F/\delta\phi(\mathbf{r})$, the ECP is an operational, coarse-grained scalar that characterizes the net mass-exchange tendency of the entire driven droplet, whose chemical-potential field is generally spatially nonuniform. Since this chemical potential decreases monotonically with droplet size, the mapping $S\leftrightarrow S^*$ serves as an effective constitutive relation: it encodes how a given structured flow modifies the effective material-exchange tendency of droplets of different sizes.
Figure~2g shows the measured $S\leftrightarrow S^*$ relations for square vortices with the same geometry but different values of $v_0$. In the absence of flow, a circular droplet would follow the trivial relation $S^*=S$. In structured vortices, the curves lie below this reference, indicating that the flow enhances interfacial curvature and raises the ECP relative to a passive droplet of the same area. More strikingly, portions of the curves have negative slope, so that increasing $S$ can decrease $S^*$ and hence increase the ECP. This inversion of the usual size--chemical-potential relation provides the basis for stable coexistence and flow-induced flux reversal.
The same constitutive relation allows us to construct phase portraits for interacting droplets. Figure~2h shows the case of two droplets in identical vortices, where arrows indicate the predicted direction of size evolution at each $(S_1,S_2)$. Simulated trajectories follow these predictions: initial conditions in region I evolve toward equal-sized droplets, region II supports stable coexistence with unequal sizes, and region III shows no stable coexistence. Details of the ECP construction and phase-portrait analysis are given in Sec.~S5.

The structure of the phase portrait follows directly from the $S\leftrightarrow S^*$ constitutive relation, viewed as an effective dynamical system for the droplet areas (see Sec.~S5. C). In particular, stable two-droplet coexistence requires at least one droplet to lie on a negative-slope branch of the $S\leftrightarrow S^*$ curve. On such a branch, a perturbation away from an equal-ECP state generates a restoring diffusive flux: a droplet that becomes too large acquires a higher ECP and loses material, whereas a droplet that becomes too small gains material. By contrast, when both droplets lie on positive-slope branches, fixed points may still exist but are unstable, as illustrated by the equal-sized fixed points in regions II and III. The phase portrait therefore provides a design principle for droplet-size control. By choosing vortex geometries that realize desired constitutive relations, one can prescribe the stable fixed points of the two-droplet dynamics. For example, increasing the flow-strength parameter $v_0$ shifts the stable coexistence sizes (Fig.~S4b), enabling the relative droplet sizes to be tuned dynamically by changing the vortex speeds. Figure~\ref{fig:2drop}i and Vid.~3 show examples of both periodic size modulation (red lines) and on-demand size adjustment (blue lines).

\section{Flow-mediated phase separation.}

Structured flows can also reshape the kinetics and morphology of phase separation. As a simple example, we initialize a nearly homogeneous mixture in a circular vortex with $\Omega(r)\sim r$. In the absence of flow, phase separation proceeds through the nucleation and Ostwald ripening of nearly circular droplets (Vid.~4a). Under circular shear, however, the early droplets are stretched by the radial variation of angular velocity after a transient nucleation stage (Fig.~\ref{fig:llps}a and Vid.~4b). Elongated droplets at similar radial positions then merge, giving rise to concentric, layered domains (Fig.~\ref{fig:llps}b). To quantify how structured flows modify coarsening, we compute the area-weighted mean droplet area, $\langle A\rangle=\sum_i A_i^2/\sum_i A_i$, where $A_i$ is the area of the $i$th identified droplet. Figure \ref{fig:llps}c shows that the ring-forming process accelerates the initial coarsening relative to the quiescent system (cyan box). At later times, the central domain gradually shrinks and disappears, followed by a sequence of similar shrinkage events until only a single domain remains. This dynamics can be viewed as Ostwald ripening constrained by circular geometry. Because the ring-like domains have weak curvature and hence only shallow chemical-potential variations, the late-stage coarsening becomes much slower than the coarsening of circular droplets without flow (magenta box in Fig.~\ref{fig:llps}c).

Structured flows can further arrest coarsening and organize phase-separated structures at larger scales. In a periodic array of identical vortices, phase separation is arrested at the single-vortex scale, and the minority phase spontaneously forms a droplet lattice whose morphology follows the imposed streamlines (Fig.~\ref{fig:llps}d and Vid.~5). Varying the vortex geometry or the mean composition produces distinct lattice patterns (Sec.~S6. A). As in the two-droplet case, the stability of these many-droplet states is tied to the negative-slope branch of the $S\leftrightarrow S^{*}$ relation, which provides a restoring flux that suppresses size differences among droplets confined in identical vortices. Arrested structures also emerge in steady disordered flows. To construct the disordered flow field, we first generate an uncorrelated vorticity field by independently sampling each grid point from a standard normal distribution, and then apply a band-pass filter in Fourier space that eventually leads to a steady disordered flow of a characteristic wavenumber $k_0$ and vorticity scale $\omega_0$. The corresponding periodic, incompressible velocity field is reconstructed from the filtered vorticity using the stream-function formulation. Phase separation in such disordered flow produces isolated or interconnected domains with characteristic sizes set by the typical vortex scale (Fig.~\ref{fig:llps}e and Vid.~6). Interestingly, neighboring droplets preferentially occupy vortices with the same sign of vorticity (Fig.~\ref{fig:llps}f and Sec. S6. B), suggesting an additional selection mechanism whose origin remains to be clarified. The periodic and disordered phase patterns remain time independent within numerical resolution over the accessible simulation time, although their asymptotic stability has not been established. It is also useful to contrast quenched disorder in the flow with quenched disorder in the free energy \cite{gyure1995scaling}. Free-energy disorder creates a heterogeneous energetic landscape that slows or pins interfaces, producing metastable or slowly aging morphologies tied to the disorder realization. Here the free energy remains uniform, while the steady disordered flow continuously advects and deforms the domains. Coarsening is therefore arrested by a driven balance between hydrodynamic transport and diffusive relaxation, yielding a nonequilibrium steady morphology whose characteristic scale is set by the flow. Removing the flow would restore ordinary coarsening.

Finally, structured vortices can control not only morphology but also the degree of phase separation. This is achieved by imposing streamlines with sharp corners, which force the interface to develop regions of high curvature. As illustrated in Fig.~\ref{fig:llps}g and Vid.~7a, these high-curvature regions raise the local chemical potential and drive partial dissolution of the droplet into the surrounding phase until a steady balance is reached. Increasing the vortex speed strengthens the alignment of the interface with the streamlines, further enhancing the curvature at the corners and reducing the steady-state droplet area (Fig.~\ref{fig:llps}h and Vid.~7b). In this way, the size of single or multiple droplets can be tuned simply by varying the flow-strength parameter $v_0$ (Fig.~\ref{fig:llps}i). More broadly, structured flow provides a means to regulate the amount of material that remains phase-separated in the driven steady state.

\section{Discussion and conclusions.}

Our results recast flow-driven LLPS as a form of hydrodynamics-mediated thermodynamic regulation. In a passive mixture, droplet fate is largely ordered by the Gibbs--Thomson relation: highly curved domains have higher chemical potential and therefore tend to dissolve into less curved ones. In a structured flow, this pathway is no longer determined by droplet size alone. The relevant entity is instead the droplet together with the flow geometry that confines and deforms it. Coherent streamlines impose persistent shape constraints, thereby changing the relation between droplet area, interfacial curvature, and effective chemical potential. This relation can be shifted, flattened, or made nonmonotonic, explaining how droplets that would be unstable at equilibrium become stable, how the inter-droplet diffusive flux can be reversed, and why extended vortex fields select finite-length-scale states instead of allowing unbounded coarsening. For clarity, the single- and two-droplet configurations are dynamically stable fixed points, whereas the many-droplet morphologies are long-lived arrested states that remain unchanged over the simulated time window. Thus, flow geometry becomes an active ingredient in the thermodynamic evolution of a phase-separating fluid.

Taken together, these results suggest a set of practical forward-design principles linking flow geometry to droplet behavior. Although they do not constitute a general inverse-design framework, they provide qualitative guidance for anticipating pinning, dissolution, mass transfer, and pattern selection, while the measured $S\leftrightarrow S^*$ relation enables quantitative predictions for interacting droplets. We summarize these principles below.
\begin{itemize}
\item \textbf{Pinning location.}
Droplets are typically trapped near the center of a closed recirculating region. More generally, a stable pinning position is one at which a small displacement produces an asymmetric interfacial deformation and a restoring diffusive flux. Thus, pinning is determined not by advection toward a stagnation point, but by the chemical-potential landscape generated as the interface samples the surrounding streamlines.

\item \textbf{Direction of mass transfer between droplets.}  
A qualitative estimate can first be made from the flow geometry. If the larger droplet is forced to align with sharper streamline features than the smaller one, its interfacial curvature and ECP are raised, making it more likely to feed the smaller droplet. If the flow imposes weaker curvature enhancement on the larger droplet, the usual Ostwald-ripening flux is expected. For a quantitative prediction, we use the measured $S\leftrightarrow S^*$ constitutive relations. Because a larger $S^*$ corresponds to a lower ECP, material flows from the droplet with smaller $S^*$ to the one with larger $S^*$; thus, the large droplet feeds the small one when $S^*_{\mathrm{large}}<S^*_{\mathrm{small}}$, and consumes it when the ordering is reversed.

\item \textbf{Sign and magnitude of the ECP shift.}
The effective chemical-potential shift is set by how the imposed flow changes the droplet’s interfacial-curvature distribution relative to its undeformed state. Its magnitude generally increases with the extent of this curvature redistribution, which is controlled by both the streamline geometry and the flow strength. To obtain an ECP that increases with droplet size, corresponding to a negative-slope branch of the $S\leftrightarrow S^*$ relation, the flow geometry must force larger droplets to align with progressively sharper portions of the streamlines. Smaller droplets then sample gentler curvature and have lower ECP, whereas larger droplets experience stronger curvature enhancement and acquire higher ECP.

\item \textbf{Dissolution versus survival.}  
Dissolution occurs when the droplet interface aligns with sharp corners of the streamlines. This alignment imposes regions of high interfacial curvature, raises the droplet’s effective chemical potential, and drives material into the surrounding phase. Whether a droplet dissolves or survives therefore depends primarily on the curvature of the streamline corners sampled by the interface and on the flow strength, which controls how strongly the interface is constrained to follow the streamlines. For a fixed flow geometry, varying the flow strength provides the principal means of promoting dissolution or permitting droplet survival and nucleation.

\item \textbf{Lattice spacing.}  
In vortex arrays, the dominant length scale is imposed by the spacing and size of the vortex cells. Coarsening is arrested when droplets become confined at this scale, producing a lattice commensurate with the underlying flow. The detailed occupation pattern is not determined by vortex spacing alone, however, and also depends on the mean composition, vortex geometry, and interactions among neighboring domains.

\end{itemize}

All results reported here are two-dimensional. The most direct experimental connection is therefore to quasi-two-dimensional geometries, such as thin microfluidic chambers and Hele--Shaw cells, where confinement suppresses out-of-plane variations and the depth-averaged dynamics can approach the description considered here~\cite{suzuki2024arresting,zhao2020review}. The underlying curvature--chemical-potential mechanism, however, is not intrinsically restricted to two dimensions. In three dimensions, structured flows can redistribute the mean curvature of a droplet surface and thereby generate chemical-potential gradients that bias diffusive mass transport~\cite{voorhees1992ostwald}. We therefore expect qualitative effects such as flow-induced migration, modified mass exchange, and droplet-size regulation to persist. The quantitative constitutive relations, stability boundaries, and collective morphologies obtained here should not, however, be transferred directly to three dimensions. Three-dimensional droplets possess additional modes of deformation and instability, including out-of-plane motion, necking, thread formation, breakup, coalescence, and Rayleigh--Plateau-type instabilities, which may compete with curvature-driven diffusion~\cite{stone1994dynamics}. Extending the present framework to fully three-dimensional structured flows will therefore require explicit treatment of surface deformation, topology changes, and hydrodynamic instabilities.

Several observations point beyond this quasi-static single-droplet picture. Droplets can oscillate spontaneously at intermediate flow speeds (Vid.~8), despite the dissipative nature of the underlying Cahn--Hilliard dynamics, suggesting feedback between advective deformation, interfacial relaxation, and diffusive equilibration. Vortex arrays also produce ordered and defective droplet patterns, including selective occupation of vortices with particular vorticity signs, which likely reflects collective coupling through the composition field. Understanding these effects will require a theory of many-droplet interactions mediated by structured chemical-potential fields.

Experimentally, the structured flows considered here can be approached using several existing flow-control strategies. Single-vortex fields, such as the circular, clover-shaped, and sharp-corner vortices used to manipulate individual droplets, can be generated in confined microfluidic geometries through patterned channel designs ~\cite{Huang2013,FernandezMateo2021}, embedded obstacles, or localized forcing elements that create recirculating flows~\cite{Stone2004}. For example, microfluidic chambers with engineered boundaries or arrays of independently driven actuators can produce stable vortical flows with tunable symmetry and circulation patterns, providing direct analogues of the vortex geometries explored in this work. Periodic vortex arrays can be realized by arranging repeated microstructures, rotating elements~\cite{Butaite2019}, or electrohydrodynamic actuators, where the spacing and symmetry of the array determine the characteristic length scale and lattice organization of the resulting phase-separated structures~\cite{Stone2004}. More complex disordered flows can be generated through randomly patterned boundary actuation~\cite{Huang2013}, acoustic streaming~\cite{Stone2004}, or active-fluid systems~\cite{Ross2019}, where the spatial correlation length of the flow can be controlled independently from its strength. Thus, although the flow fields considered here are idealized target geometries, existing experimental platforms already provide routes to realizing comparable streamline topologies, including closed vortices, vortex arrays, and spatially heterogeneous flows.

Looking ahead, time-dependent forcing and inverse design provide natural extensions. At present, our framework maps a prescribed flow to the resulting droplet dynamics, but does not solve the inverse problem of constructing a flow for a desired state. The measured $S\leftrightarrow S^*$ constitutive relation provides a first step by predicting stable droplet sizes within a calibrated family of vortices. A more general design map would need to connect the effective chemical potential to droplet geometry, flow strength, interfacial properties, and streamline topology. Such a framework could ultimately enable the systematic construction of flow fields that realize prescribed droplet sizes, morphologies, or dynamical pathways.
More broadly, structured hydrodynamic driving provides a framework for examining how nonequilibrium fields modify thermodynamic relaxation, stability, and pattern selection in extended soft-matter systems.

\bibliographystyle{apsrev4-1}
\bibliography{flow_regulate_LLPS}

\section*{Acknowledgements}
ZY was supported by the National Natural Science Foundation of China No. 12374219, the National Key Research and Development Program of China (No. 2023YFA1407500), the 111 project (B16029). QZ was supported by the Scientific Research Funding of Jiangsu University of Science and Technology (No.1052932204).

\section*{Conflicts of interest}
There are no conflicts of interest.

\end{document}


\title{Supplemental materials to Structured flows regulate droplet dynamics in phase-separating fluids}

\author{Yulin Li}
\thanks{These authors contributed equally}
\affiliation{Fujian Provincial Key Laboratory for Soft Functional Materials Research, Research Institute for Biomimetics and Soft Matter, Department of Physics, Xiamen University, Xiamen, Fujian 361005, China}

\author{Tong Zhou} 
\thanks{These authors contributed equally}
\affiliation{Fujian Provincial Key Laboratory for Soft Functional Materials Research, Research Institute for Biomimetics and Soft Matter, Department of Physics, Xiamen University, Xiamen, Fujian 361005, China}
 
\author{Yanyu Li}
\thanks{These authors contributed equally}
\affiliation{Fujian Provincial Key Laboratory for Soft Functional Materials Research, Research Institute for Biomimetics and Soft Matter, Department of Physics, Xiamen University, Xiamen, Fujian 361005, China}

\author{Qi Zhang}
\affiliation{School of Science, Jiangsu University of Science and Technology, Zhenjiang, Jiangsu 212100, China}

\author{Zhihong You}
\thanks{Corresponding author: zhyou@xmu.edu.cn}
\affiliation{Fujian Provincial Key Laboratory for Soft Functional Materials Research, Research Institute for Biomimetics and Soft Matter, Department of Physics, Xiamen University, Xiamen, Fujian 361005, China}
	
\date{\today}
	
\maketitle

\section{Numerical implementations}
This section describes the numerical implementation used in our simulations, including the solver for Eq.~(1) in the main text and the construction of the prescribed structured flow fields.

We first construct incompressible flow fields using the vorticity--stream function method~\cite{barragy1997stream}. Specifically, we begin by prescribing a target velocity field $\mathbf{v}^{\text{tar}}$ with the desired spatial structure, which in general is not exactly incompressible. We then compute  its vorticity,
$\omega=\hat{\mathbf{z}}\cdot\nabla\times\mathbf{v}^{\text{tar}}$, and obtain the corresponding streamfunction $\psi$ by solving the Poisson equation $\nabla^2\psi=-\omega$. The incompressible velocity field is then reconstructed as
$v_x=\partial_y\psi$ and $v_y=-\partial_x\psi$. This procedure yields a flow field $\mathbf{v}=(v_x,v_y)$ that closely preserves the intended structure of $\mathbf{v}^{\text{tar}}$ while satisfying $\nabla\cdot\mathbf{v}=0$. The specific target fields $\mathbf{v}^{\text{tar}}$ used for each flow geometry are given in Sec.~S7. We represent $\mathbf{v}^{\text{tar}}=v_0\mathbf{\hat{v}}^{\text{tar}}$ where $v_0$ is the velocity amplitude parameter and $\mathbf{\hat{v}}^{\text{tar}}$ is a normalized velocity field satisfying $\max_{\mathbf{r}}|\hat{\mathbf{v}}^{\mathrm{tar}}(\mathbf{r})|=1$.

Given the prescribed flow field, Eq.~(1) is solved using a finite-difference method in a square domain of size $L_x\times L_y$. Periodic boundary conditions are applied in both spatial directions to the phase field $\phi$, chemical potential $\mu$, and prescribed velocity field $\mathbf{v}$. Space is discretized on a uniform square grid with spacing $h$, and spatial derivatives are evaluated using a fourth-order central-difference scheme. The characteristic diffuse-interface length is $\ell=\sqrt{k/a}$. We use the nondimensional units defined in the main text and set $a=k=1$, such that $\ell=1$. The mobility is taken to be spatially uniform and composition independent, with $M=1$. Time integration is performed using a fixed-step Runge--Kutta--Chebyshev scheme, which provides an extended stability range for the stiff Cahn--Hilliard dynamics~\cite{verwer1990convergence}. Unless otherwise specified, we use $L_x=L_y=128$, $h=1$, and $\Delta t=0.1$.

The initial phase field $\phi$ is chosen either as a single circular droplet or as a nearly homogeneous mixture with a prescribed mean composition. For a single circular droplet, we define the initial phase-field profile as $\phi_c(\mathbf{r};\mathbf{r}_c,R_0)=[1+\tanh(R_0-|\mathbf{r}-\mathbf{r}_c|)]/2$, where $\mathbf{r}_c$ and $R_0$ denote the droplet center and radius, respectively. For a two-droplet configuration, the initial condition is $\phi(\mathbf{r},0)=\phi_c(\mathbf{r};\mathbf{r}_{c,1},R_{0,1})+\phi_c(\mathbf{r};\mathbf{r}_{c,2},R_{0,2})$, where $\mathbf{r}_{c,j}$ and $R_{0,j}$ are the center and radius of droplet $j$, respectively. This gives a diffuse interface of width of order unity. For simulations initialized from a homogeneous state, we set $\phi(\mathbf{r},0)=\phi_0+\delta\phi(\mathbf{r})$, where $\phi_0$ is the mean composition of the mixture and $\delta\phi$ is a small random perturbation.


\section{Discussions on the Imposed-Flow Approximation}

In the manuscript, we treat the flow as externally imposed and neglect hydrodynamic back-coupling from the concentration field. In this section, we discuss the conditions under which this approximation is justified and quantify its validity using direct simulations of the full Cahn--Hilliard--Navier--Stokes (CHNS) model.

In general, when an interface is deformed away from equilibrium, the resulting imbalance of interfacial tension generates capillary stresses that drive a flow tending to restore the interface. The imposed-flow approximation therefore corresponds to a regime of weak back-coupling, in which velocity variations induced by capillary stresses are small compared to those produced by the external driving that maintains the flow patterns of interest. The latter can be represented generically by a linear relaxation term $\mathbf{{f}}=\tau_f^{-1}(\mathbf{v}^{\text{tar}}-\mathbf{v})$ in the Navier–Stokes equation, without specifying the microscopic origin of the driving. Here, $\mathbf{v}^{\text{tar}}$ denotes the target flow field and $\tau_f$ the characteristic relaxation time toward $\mathbf{v}^{\text{tar}}$. By contrast, interfacial tension drives flow on a distinct timescale: for a strongly deformed interface, the fastest capillary response is associated with the smallest local radius of curvature,
\begin{equation}
    R_\kappa=
    \frac{1}{|\kappa|_{max}},
    \qquad
    \kappa=
    \nabla\cdot
    \left(\frac{\nabla\phi}{|\nabla\phi|}\right).
\end{equation}
In the viscous-capillary estimate, the corresponding relaxation time is
\begin{equation}
    \tau_c\sim\frac{\eta R_\kappa}{\gamma},
\end{equation}
where $\eta$ is the dynamic viscosity and $\gamma$ is the two-dimensional interfacial
tension. Thus, the relevant comparison is between the external relaxation time $\tau_f$ and the local capillary-response time $\tau_c$.
When $\tau_f\ll \tau_c$, the velocity field rapidly converges to $\mathbf{v}^{\text{tar}}$, rendering capillary-induced flows negligible and justifying the decoupling of the velocity from the composition field.

\begin{figure}[t]
    \centering
    \includegraphics[width=\textwidth]{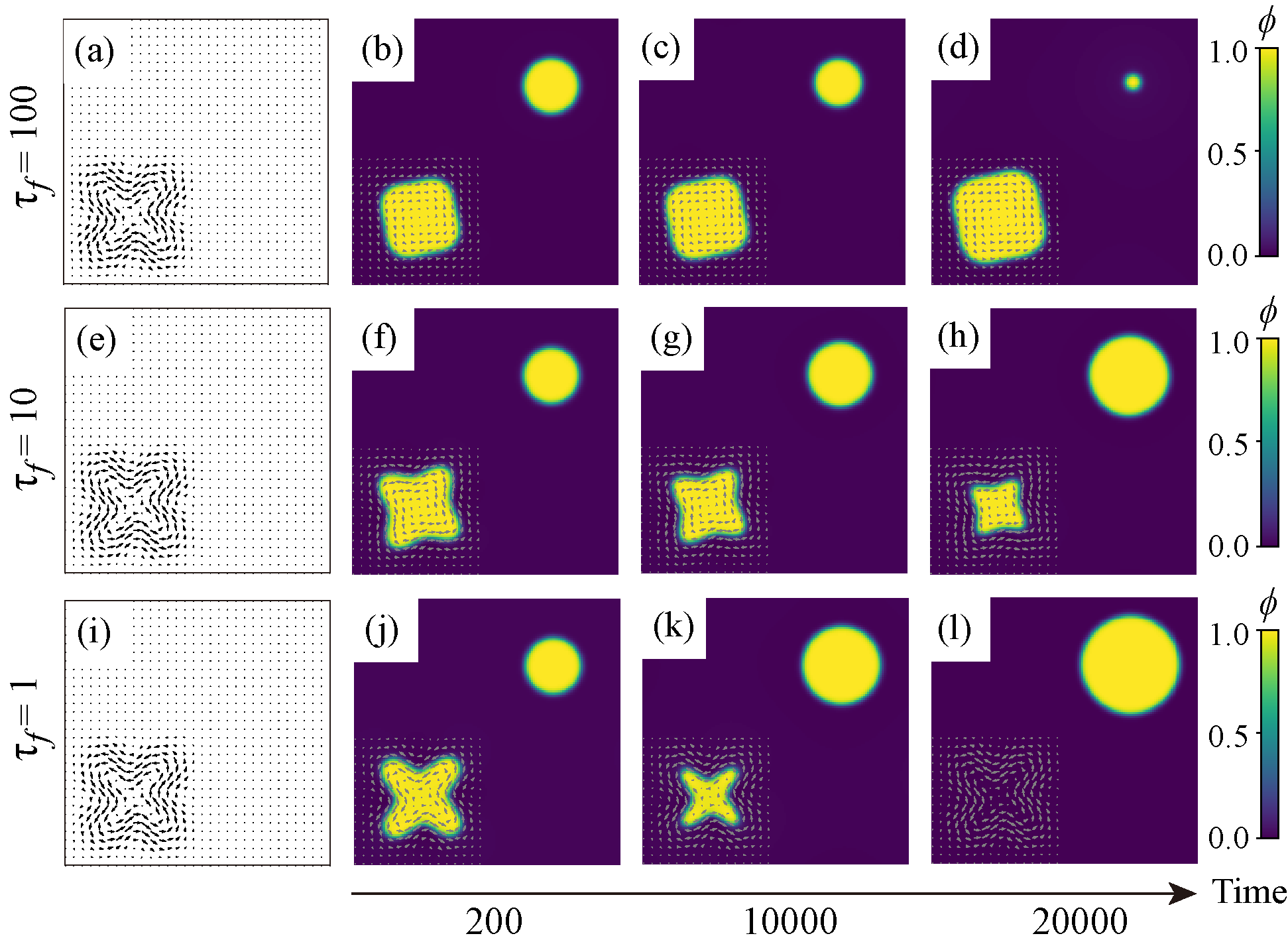}    
    \caption{
   Flow and composition fields for different values of $\tau_f$. (a), (e), and (i): Prescribed target flow field $\mathbf{v}^{tar}$. (b)-(d): Two-droplet dynamics for $\tau_f=100$. (f)-(h): Two-droplet dynamics for $\tau_f=10$. (j)-(l): Two-droplet dynamics for  $\tau_f=1$. }
    \label{figR:diff_tau}
\end{figure}
To validate this argument, we perform simulations of the full Cahn–Hilliard–Navier–Stokes (CHNS) equations, explicitly evolving the velocity field and including interfacial effects through the divergence of the Korteweg stress $\mu\nabla\phi$:
\begin{subequations}
\label{eq:system}
\begin{align}
    \label{eq:phi_sub}
    \partial_t \phi + \textbf{v} \cdot \nabla \phi &= M \nabla^2 \mu, \\
    \label{eq:flow_sub}
    \rho (\partial_t \textbf{v} + \textbf{v} \cdot \nabla \textbf{v}) &= -\nabla p + \eta \nabla^2 \textbf{v} + \mu \nabla \phi + \rho\textbf{\textit{f}}.
\end{align}
\end{subequations}
The external driving $\mathbf{f}$ is implemented via the same linear relaxation term introduced above. Incompressibility is enforced through the condition $\nabla\cdot\mathbf{v}=0$. The velocity field is initialized as $\mathbf{v}(\mathbf{r},0)=\mathbf{v}^{\text{tar}}(\mathbf{r})$, corresponding to an externally established flow before the phase-field evolution begins. In the simulations, we set $\rho=1$ and $\eta=1$. The term $\mu\nabla\phi$ represents the Korteweg interfacial force, through which the evolving phase field can, in principle, feed back onto the flow. The coupled CHNS simulations are characterized by $Re=\rho U L_c/\eta\simeq16$--$96$ and $Ca=\eta U/\gamma\simeq8.5$--$25.5$, where $U$, $L_c$, and $\gamma$ are the characteristic velocity, droplet size, and interfacial tension, respectively. The parameters of the Cahn-Hilliard equation and the simulation setup are the same as those in Sec. S1.

We use flow-induced flux reversal (Figs. 2a-c in the main text) as a representative example to evaluate the role of $\tau_f$. In these simulations, for the free-energy density $f(\phi)=a\phi^2(1-\phi)^2/4$, the equilibrium interfacial tension is
\begin{equation}
    \gamma
    =\int_0^1\sqrt{2k f(\phi)}\,d\phi
    =\frac{\sqrt{ak}}{6\sqrt{2}}.
\end{equation}
With $a=k=1$, this gives $\gamma\simeq0.118$. From the $\phi=1/2$ contour of the strongly deformed droplet, we obtain a minimum local radius of curvature $R_\kappa\simeq2.5$. For $\eta=1$,
the corresponding local viscocapillary time is therefore
\begin{equation}
    \tau_c
    \sim\frac{\eta R_\kappa}{\gamma}
    \simeq21.
\end{equation}
This value is an order-of-magnitude estimate for the relaxation of the most strongly curved regions, rather than the relaxation time of the entire droplet. The estimate $\tau_c=\eta R_\kappa/\gamma$ shows explicitly that the crossover depends on viscosity. The present comparison uses $\eta=1$; therefore, the resulting validity range should be understood for this viscosity rather than as a universal threshold in $\tau_f$.
Figure~\ref{figR:diff_tau} shows the evolution of the flow and composition fields for different values of $\tau_f$ (second to fourth columns), alongside the target flow $\mathbf{v}^{\text{tar}}$ (first column). For $\tau_f\gg\tau_c$ (top row), the external driving is weak and interfacial stresses produce a noticeable back-coupling on the flow (Figs. \ref{figR:diff_tau}a and \ref{figR:diff_tau}b), leading to conventional Ostwald ripening. When $\tau_f\approx \tau_c$ (middle row), external driving and interfacial effects are comparable. Interfacial feedback remains visible in the flow (Figs. \ref{figR:diff_tau}e and \ref{figR:diff_tau}f), yet flow-induced flux reversal can already occur. Finally, in the regime $\tau_f\ll\tau_c$ (bottom row), the external driving dominates, interfacial contributions to the flow become negligible (Figs.~\ref{figR:diff_tau}i and \ref{figR:diff_tau}j), and flow-induced flux reversal proceeds as described in the main text.

\begin{figure}[h]
\centering
\includegraphics[width=0.6\textwidth]{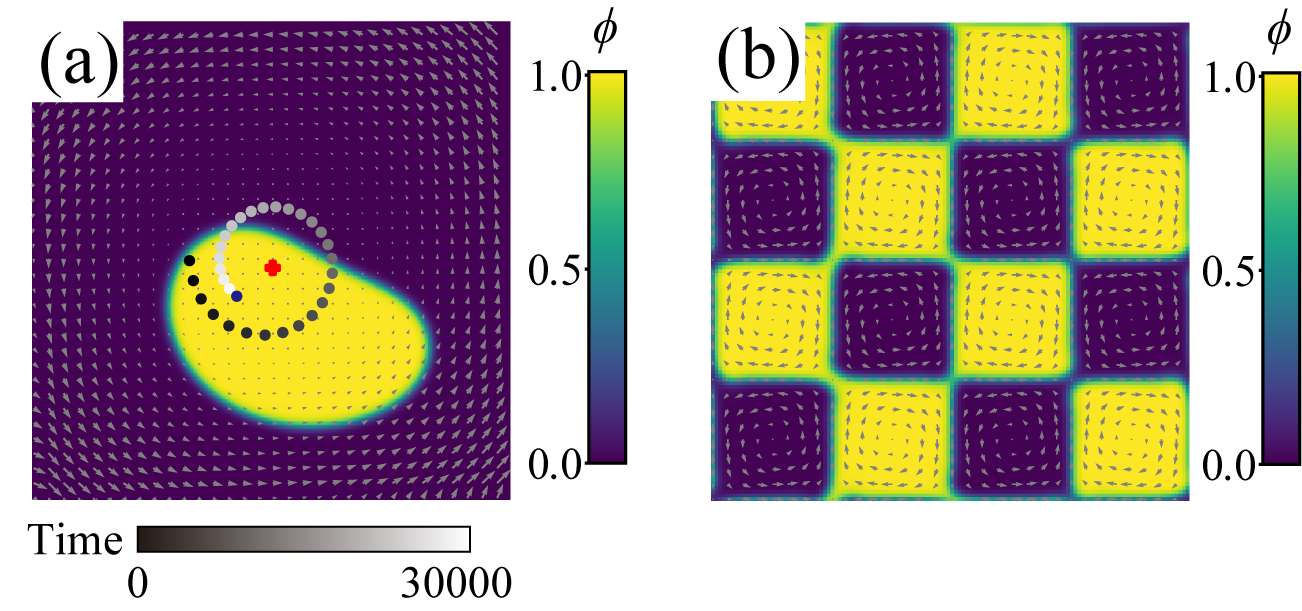}
\caption{Two-way-coupled hydrodynamic simulations of droplet pinning and lattice formation with $\tau_f=1$. (a) Droplet pinning under the same single-vortex geometry as that used in Fig. 1a. (b) Formation of an alternating lattice-like phase pattern under the same periodic vortex array as that used in Fig. S5f.}
\label{figS:hydro2}
\end{figure}

To examine more directly whether hydrodynamic back-coupling affects the pinning and lattice-formation behaviors discussed in the main text, we performed two additional CHNS simulations using the same relaxation time, $\tau_f=1$, and the same material parameter (Fig.~\ref{figS:hydro2}).

In the single-vortex geometry, the droplet deforms under the local shear, but its centroid remains confined around the vortex center rather than escaping from the trapping region (Fig.~\ref{figS:hydro2}a). Thus, the pinning behavior found in the prescribed-flow model persists when the droplet is allowed to feed back onto the velocity field through the interfacial force. In the periodic vortex-array geometry, the phase-separated domains organize into similar alternating lattice-like pattern as in the one-way-coupled simulations (Fig.~\ref{figS:hydro2}b). The velocity field is locally perturbed near the interfaces, but its principal vortex topology and the resulting spatial organization are retained.

Taken together, the three representative two-way-coupled cases show that the flow-induced flux reversal, droplet pinning, and lattice formation remain qualitatively unchanged in the externally maintained-flow regime considered here. Hydrodynamic back-coupling may modify quantitative details, such as the transient trajectory, relaxation time, interface shape, and precise droplet position, but it does not alter the underlying mechanisms or the principal conclusions.

In this weak-feedback regime, the CHNS dynamics approaches the advective Cahn--Hilliard equation with the prescribed target velocity field,
\begin{equation}
\label{eq:phi1}
\partial_t\phi+
\mathbf{v}^{\text{tar}}\cdot\nabla\phi
=
M\nabla^2\mu.
\end{equation}
These results therefore support the one-way-coupling approximation adopted in the main text for the parameter regime studied here.

In these tests, the two phases are taken to have matched density and viscosity, so that the simulations specifically examine hydrodynamic feedback generated by interfacial stresses. The conclusions therefore apply to an externally maintained, density- and viscosity-matched regime, and do not imply that feedback remains weak for high droplet loading, strong density or viscosity contrasts, or weak external forcing. Under such conditions, the flow field and the resulting morphology may become strongly coupled and system dependent.

\section{Extracting the area and position of droplets}

Since the mean composition considered in this work does not exceed $\phi=0.5$, the minority phase, corresponding to $\phi=1$, typically forms droplets embedded in the majority phase with $\phi=0$ (see, for example, Fig.~1a in the main text). We identify droplets as connected components of grid points satisfying $\phi\ge 0.5$. The area \(A_i\) of the \(i\)-th droplet is defined as the total area of the corresponding connected component, where \(i=1,\ldots,N_d\) indexes the identified droplets and \(N_d\) is the total number of droplets.

For systems containing multiple droplets, we characterize the typical droplet size by an area-weighted mean, $\langle A\rangle = \sum_i A_i^{2}/\sum_i A_i= \sum_i \omega_i A_i/\sum_i \omega_i$, with $\omega_i=A_i$. This definition weights each droplet in proportion to its area fraction. It is therefore less sensitive to small transient droplets and better reflects the characteristic length scale of the morphology during coarsening, where the late-time dynamics is dominated by the larger domains.

For simulations containing a single droplet, we compute its center of mass as $\mathbf{R}_c=(\mathbf{X}_c,\mathbf{Y}_c)$, $\mathbf{X}_c=\sum_j \mathbf{x}_j\phi_j/M$, $\mathbf{Y}_c=\sum_j \mathbf{y}_j\phi_j/M$, where the sums run over all grid points, $\phi_j$ is the phase-field value at grid point $j$, and $M=\sum_j \phi_j$ is the total amount of the minority phase.

\section{Alignment of droplet interface to the flow streamlines}

\begin{figure}[h]
    \centering
    \includegraphics[width=0.8\textwidth]{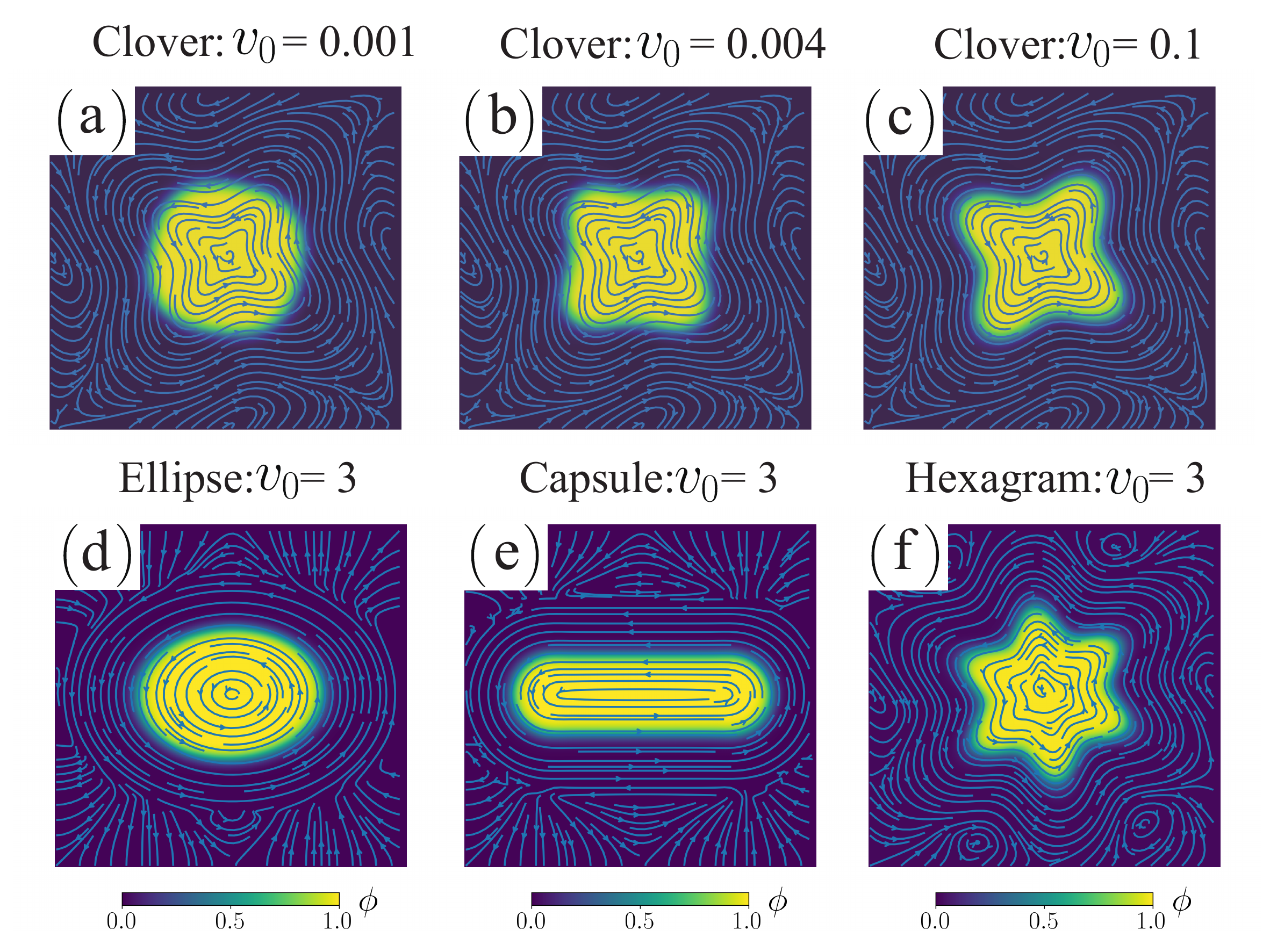}    
    \caption{Alignment of droplet interface to the flow streamline. (a-c) Steady shape of droplets in clover-shaped vortices with the same structure but different flow strengths \(v_0\). (d-e) Steady shape of droplets in different structured vortices. The detailed vortex structures can be found in Sec. S7D.}
    \label{figS:align}
\end{figure}

When placed in a structured flow, a droplet interface tends to align with the local streamlines. The degree of alignment is controlled by the velocity gradients in the flow. As shown in Figs.~\ref{figS:align}a--\ref{figS:align}c, increasing the velocity gradient leads to stronger interfacial alignment. This behavior reflects a competition between advective deformation and interfacial tension. Advection tends to rotate and stretch any interface segment that is not locally parallel to a streamline, whereas interfacial tension tends to restore the droplet toward a circular shape. Parallel alignment corresponds to the configuration in which the local shear-induced rotation of the interface is minimized. Thus, increasing the flow strength \(v_0\) strengthens the advective contribution relative to interfacial relaxation, producing closer alignment with the streamlines (Figs.~\ref{figS:align}a--\ref{figS:align}c). Figures~\ref{figS:align}d--\ref{figS:align}f illustrate the steady droplet shapes selected by several different flow geometries.

\section{Constitutive relation and phase portrait}

This section develops the effective constitutive relation and phase-portrait construction for droplets in structured vortices, which together provide a quantitative framework for predicting droplet-size evolution.

\subsection{Constitutive relation}

To predict the size evolution of droplets in structured vortices, we need to determine how their effective chemical potential (ECP) depends on droplet size. This relation plays a role analogous to an equation of state: for a fixed vortex geometry, the droplet area serves as a state variable, while the ECP determines the direction of diffusive mass exchange. We therefore refer to the ECP--area relation as the constitutive relation of a droplet in a structured vortex.

Unlike the pressure of an ideal gas, however, the ECP cannot be read off directly from a local field. In a deformed droplet, the chemical potential is spatially inhomogeneous (see, e.g., Figs.~1b and 2b), making a direct definition ambiguous. We therefore determine the ECP operationally. For a given vortex, we place a droplet of area $S$ inside the flow and a circular reference droplet in a region where the velocity is negligible, as illustrated in Fig.~2a of the main text. By varying the area of the reference droplet, we identify a value $S^*$ at which the net diffusive flux between the two droplets vanishes. The ECP of the deformed droplet is then defined as the equilibrium chemical potential of a circular droplet of area $S^*$. Repeating this procedure for different values of $S$ gives a mapping $S\leftrightarrow S^*$. Since the chemical potential of a circular droplet decreases monotonically with its area, the precise numerical value of the ECP is not needed: for two droplets in structured vortices, the droplet with larger $S^*$ has lower ECP and therefore tends to grow. Thus, the $S\leftrightarrow S^*$ mapping provides the constitutive relation needed to predict droplet-size evolution. \red{Because the ECP is determined operationally through coexistence with a reference droplet, the measured ($S^*$) may retain some dependence on the separation between the two droplets, owing to finite-range interactions of their chemical-potential fields and finite-domain effects. We therefore place the reference droplet in a flow-free region as far as possible from the vortex-confined droplet. The resulting ($S\leftrightarrow S^*$) relation should be understood as an effective constitutive relation for this fixed measurement geometry.}

We now discuss the shape of the $S\leftrightarrow S^*$ curves shown in Fig.~2g of the main text. For a circular droplet in the absence of flow, the mapping is simply $S^*=S$. In structured vortices, by contrast, the curves lie well below this reference line. This indicates that the flow reduces the equivalent reference size $S^*$ and hence raises the ECP of the droplet, primarily by imposing additional interfacial curvature. In some ranges, the curves even acquire a negative slope: increasing the droplet area decreases $S^*$ and therefore increases the ECP. This behavior is opposite to that of passive equilibrium droplets and provides the basis for stable coexistence, as discussed below. Such a negative-slope branch is not generic. For example, a circular vortex centered on a droplet does not appreciably change its ECP. To generate a negative slope, the vortex geometry must force larger droplets to sample regions of sharper curvature, so that the mean interfacial curvature increases with droplet size. This illustrates how structured vortices can reshape the effective thermodynamic response of droplets.

\subsection{Phase portrait}
A phase portrait is a standard tool in dynamical systems for visualizing the global structure of trajectories in phase space~\cite{butkovskiy2012phase}. Here we use this framework to characterize droplet-size evolution in structured vortices.

As discussed above, the area of each droplet can be treated as a state variable. A two-droplet system is therefore described by the pair $(S_1,S_2)$, where $S_1$ and $S_2$ are the areas of droplets 1 and 2. Its time evolution corresponds to a trajectory in the two-dimensional phase space spanned by these variables, as shown by the dotted curves in Fig.~2h of the main text. Rather than plotting many individual trajectories, we construct a phase-space vector field that indicates the instantaneous direction of droplet-size evolution at each point $(S_1,S_2)$. Assuming conservation of the total droplet area, $S_1+S_2=S_0$, trajectories are confined to lines of constant $S_0$, and the corresponding phase velocities are parallel to these lines. The direction of motion is determined from the constitutive relations by comparing the values of $S^*$ associated with $S_1$ and $S_2$. Since a larger $S^*$ corresponds to a lower effective chemical potential, the droplet with larger $S^*$ gains material, while the other loses material. Applying this rule throughout phase space gives the phase portrait shown in Fig.~2h.

\subsection{Discussions on the structure of phase portrait}

As discussed in the main text, the phase portrait in Fig.~2h immediately reveals three regimes of two-droplet dynamics. In region I, all arrows point toward the equal-size line $S_1=S_2$, indicating that the two droplets exchange mass until their areas become equal. In dynamical-systems terms, the equal-size states in this region are stable fixed points. In region III, by contrast, the arrows point away from $S_1=S_2$, indicating that the equal-size states are unstable: any small difference in droplet size is amplified until the smaller droplet is consumed. Between these two regimes, region II contains unstable equal-size fixed points together with two branches of stable unequal-size fixed points, shown by the yellow solid lines in Fig.~2h. This structure is characteristic of a supercritical bifurcation.

The phase portrait can be understood from the stability of fixed points determined by the $S\leftrightarrow S^*$ constitutive relation. Fixed points correspond to states in which the two droplets have equal effective chemical potentials, or equivalently equal values of $S^*$. For two droplets in identical vortices, the equal-size states $S_1=S_2$ are therefore always fixed points. Because the $S\leftrightarrow S^*$ curve is nonmonotonic (Fig.~2g), additional unequal-size fixed points can also occur when $S^*(S_1)=S^*(S_2)$ with the two droplets lying on different branches of the constitutive relation. These unequal-size fixed points are shown as the orange solid lines in Fig.~2h.

\begin{figure}[h]
    \centering
    \includegraphics[width=1.0\textwidth]{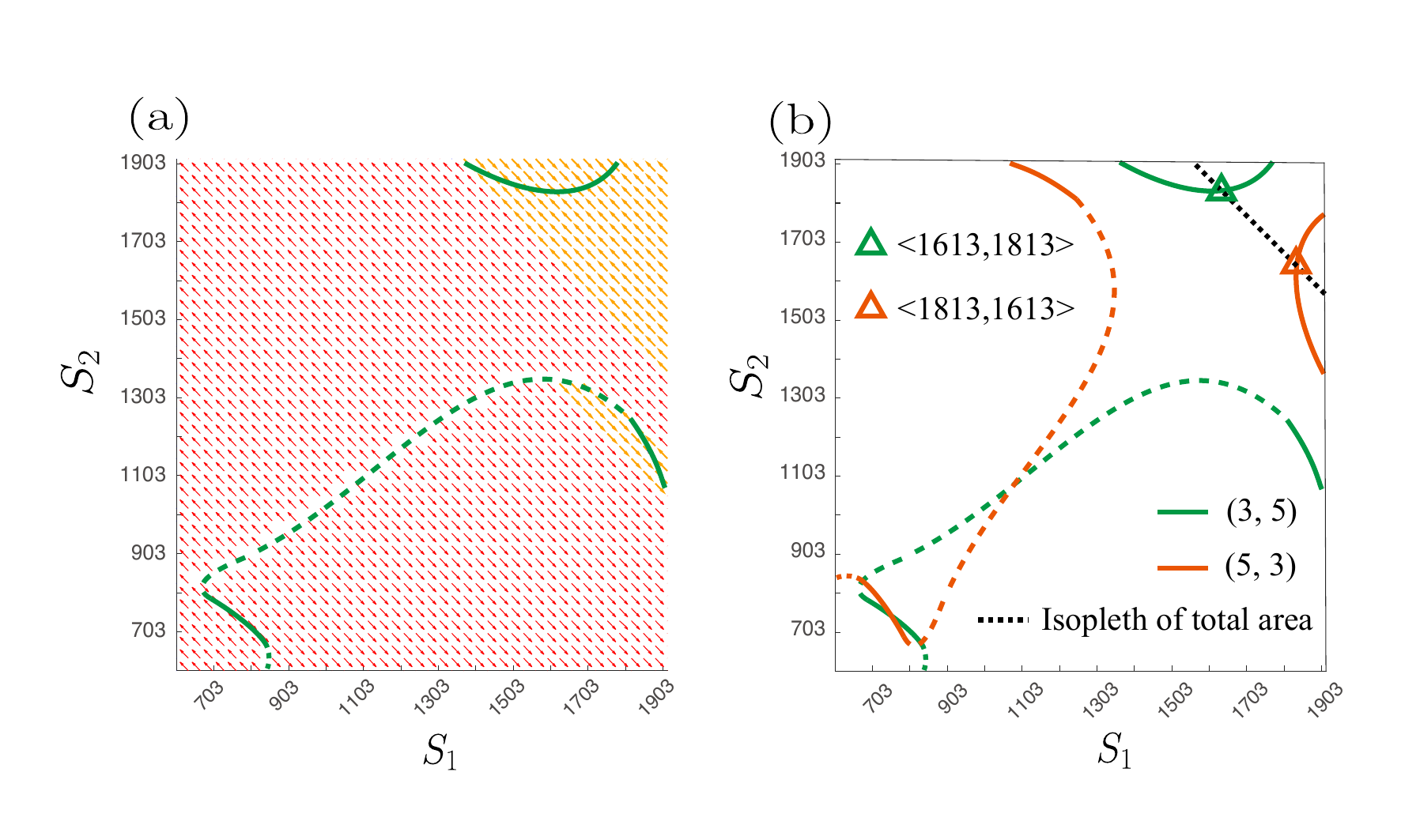}    
    \caption{(a) Portrait of two droplets placed in square vortices with different speeds: $v_{01}=3$ and $v_{02}=5$. The solid and dashed lines represent stable and unstable fixed points, respectively. (b) Fixed points of droplets at $(v_{01}=5,v_{02}=3)$ (orange lines) and $(v_{01}=3,v_{02}=5)$ (green lines).}
    \label{figS:PhasPort}
\end{figure}

The stability of the equal-size states follows directly from the local slope of the $S\leftrightarrow S^*$ curve. Consider a small perturbation around $S_1=S_2$, such that the droplet areas become $S_1-\delta S$ and $S_2+\delta S$ with $\delta S>0$. On a positive-slope branch, the ECP decreases with increasing area. The smaller droplet therefore has a higher ECP than the larger one, producing a diffusive flux that further increases the size difference; the equal-size fixed point is unstable. On a negative-slope branch, the response is reversed: the larger droplet has the higher ECP and loses material, while the smaller droplet gains material. The perturbation is therefore suppressed, making the equal-size fixed point stable. These two cases correspond to the phase velocities observed in regions III/II and region I, respectively.

The unequal-size fixed points can be analyzed in the same way. Consider a fixed point satisfying $S^*(S_1)=S^*(S_2)$ with $S_2>S_1$, and perturb it to $S_1-\delta S$ and $S_2+\delta S$. To linear order, $S^*(S_1-\delta S)\approx S^*(S_1)-\delta S\frac{dS^*(S_1)}{dS_1}$, and $S^*(S_2+\delta S)\approx S^*(S_2)+\delta S\frac{dS^*(S_2)}{dS_2}$. Thus, $\Delta S^*\equiv S^*(S_1-\delta S)-S^*(S_2+\delta S) \approx -\delta S\left[\frac{dS^*(S_1)}{dS_1}+\frac{dS^*(S_2)}{dS_2}\right]$. For the unequal-size branch in Fig.~2h, the smaller droplet lies on a positive-slope branch and the larger droplet on a negative-slope branch. Stability requires the perturbation to generate a restoring flux, which occurs when the negative slope at $S_2$ is sufficiently steep, namely $\left|\frac{dS^*(S_2)}{dS_2}\right|>\left|\frac{dS^*(S_1)}{dS_1}\right|$. This condition is satisfied for the constitutive relation shown in Fig.~2g, explaining why the unequal-size branches in Fig.~2h are stable.

The same construction applies when the two droplets are placed in vortices with different structures or speeds. By changing the vortex speed while keeping the overall geometry fixed, one shifts the location of the stable fixed points, causing the droplet sizes to redistribute toward the new steady state. This provides a mechanism for dynamic size control. For example, the red curves in Fig.~2i are obtained by switching the vortex speeds $(v_{01},v_{02})$ between $(3,5)$ and $(5,3)$, causing the droplet sizes to periodically exchange between $(1613,1813)$ and $(1813,1613)$, as indicated by the green and orange triangles in Fig.~\ref{figS:PhasPort}b.

\section{Phase separation driven by vortex array and disordered flow}

This section provides extra results on phase separation driven by vortex arrays and disordered flows.

\subsection{Vortex array}
\begin{figure}[b]
    \centering
    \includegraphics[width=0.7\textwidth]{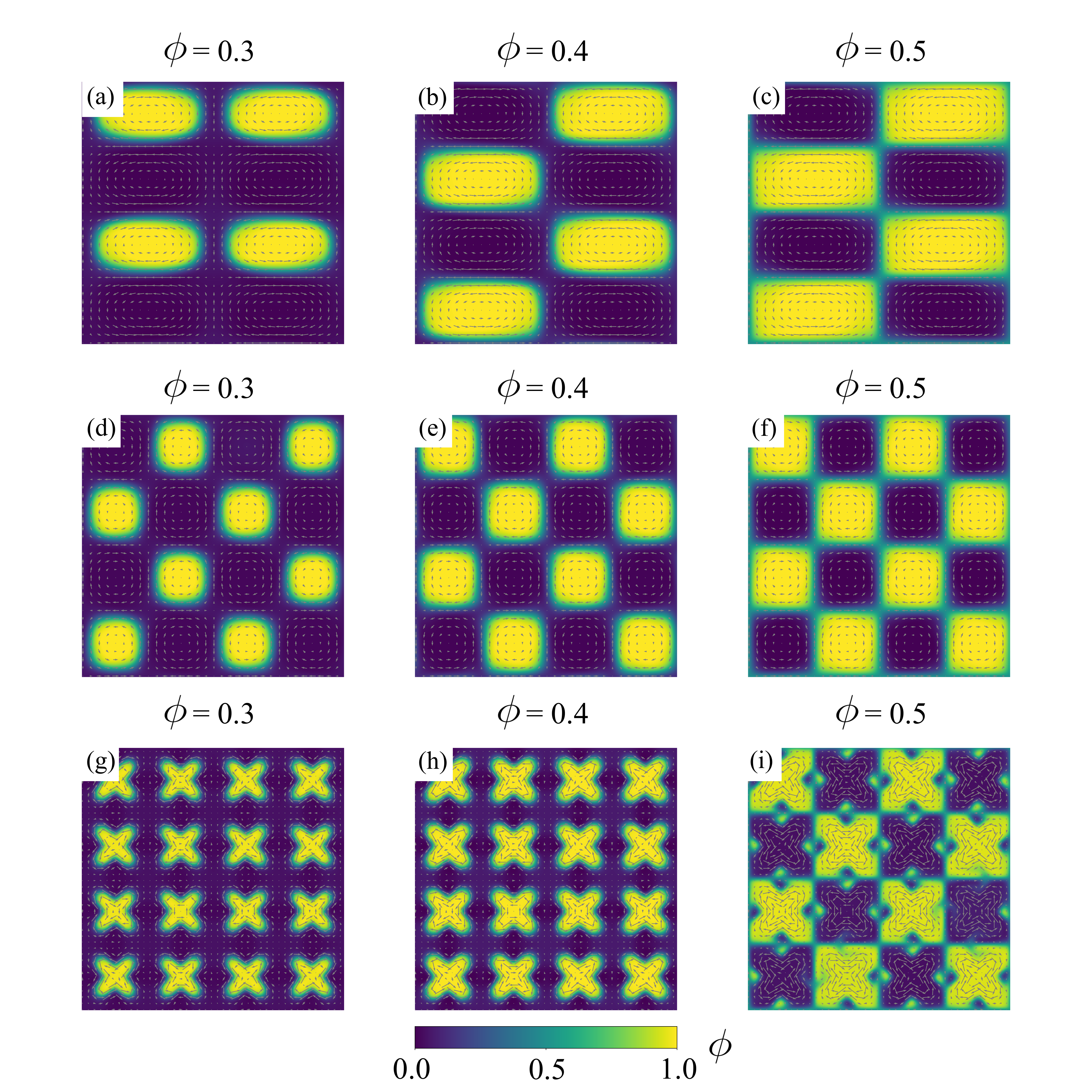}    
    \caption{Phase separation in vortex arrays with different initial compositions and flow structures.}
    \label{figS:lattice}
\end{figure}

As discussed in the main text, vortex-array-driven phase separation can produce periodic droplet lattices. The resulting lattice structure depends on both the mean composition $\phi_0$ and the geometry of the underlying vortices. Figure~\ref{figS:lattice} shows representative steady states under different conditions. In many cases, droplets occupy only a subset of the vortices, often those with the same sign of vorticity (Figs.~\ref{figS:lattice}b--\ref{figS:lattice}f). Exceptions can also occur, as in Fig.~\ref{figS:lattice}a, where droplets appear in vortices with opposite vorticity signs. Varying $\phi_0$ typically changes the droplet sizes rather than the number of occupied vortices. For vortex geometries with concave streamline segments, droplets may occupy only part of each vortex cell instead of filling the entire region (Figs.~\ref{figS:lattice}g--\ref{figS:lattice}h). The detailed selection rules underlying these lattice structures are beyond the scope of this work.

\subsection{Disordered flow}

\begin{figure}[b]
    \centering
    \includegraphics[width=0.7\textwidth]{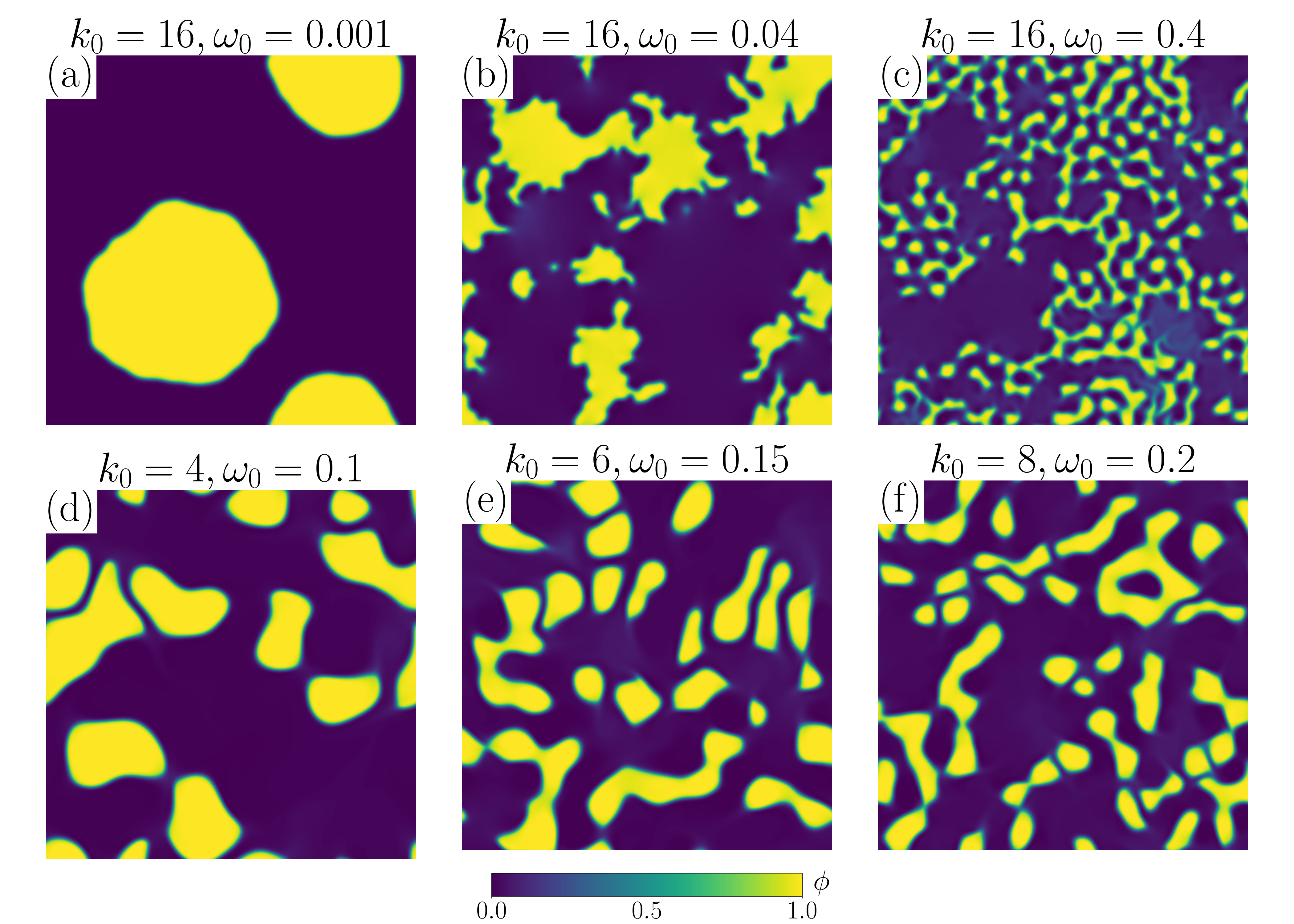}    
    \caption{Phase separation in disordered flow under different conditions.}
    \label{figS:disorder}
\end{figure}

We next examine phase separation driven by steady flows with quenched disorder. To this end, we generate random incompressible velocity fields whose characteristic length scale is controlled by a typical wavenumber $k_0$, and whose strength is set by a characteristic vorticity $\omega_0$; details are given in Sec.~S7.C. Figure~\ref{figS:disorder} shows representative \red{arrested morphologies} obtained for different flow strengths and length scales. When the flow is sufficiently strong, coarsening is arrested and the phase field reaches a disordered steady morphology. The characteristic domain size is then set by the length scale of the underlying flow.

\section{Generation of different structured flow fields}

This section collects the specific numerical protocols to generate different types of structured flows used in this article. As mentioned in Sec. S1, the vortex-stream function method is used to generate the incompressible structured flow. Before that, we need to prescribe a velocity field that is simpler but does not necessarily satisfy the condition of incompressibility.

\subsection{Circular vortices in Fig. 1}

To generate a circular vortex, we first define a few parameters: the position of the vortex center $\mathbf{r}_{\mathrm{c}}$, the characteristic velocity $v_0$, and the radial normalization length $r_{\mathrm{norm}}$. With these at hand, a circular target velocity field $\mathbf{v}^{\mathrm{tar}}(\mathbf{r})=[v_x^{\mathrm{tar}}(\mathbf{r}),v_y^{\mathrm{tar}}(\mathbf{r})]$ can be generated using $v_x^{\mathrm{tar}}(\mathbf{r})=-\frac{\Delta y}{\Delta r}\times\frac{v_0}{2}(\Delta r/r_{\mathrm{norm}})^n$ and $v_y^{\mathrm{tar}}(\mathbf{r})=\frac{\Delta x}{\Delta r}\times\frac{v_0}{2}(\Delta r/r_{\mathrm{norm}})^n$, with the velocity set to zero at $\Delta r=0$. Here, $\Delta r=|\mathbf{r}-\mathbf{r}_{\mathrm{c}}|$ is the distance to the vortex center, and $\Delta x=(\mathbf{r}-\mathbf{r}_{\mathrm{c}})_x$ and $\Delta y=(\mathbf{r}-\mathbf{r}_{\mathrm{c}})_y$. The parameter $n$ determines the radial dependence of the velocity and angular velocity. In Fig.~1, we set $L_x=L_y=128$, $\mathbf{r}_{\mathrm{c}}=[L_x/2,L_y/2]$, and $r_{\mathrm{norm}}=L_x-10=118$. The values of $v_0$ and $n$ differ for different types of flow:
\begin{itemize}
\item Uniform angular velocity: $v_0 = 0.1$ and $n = 1$, giving $\Omega(\Delta r)=4.2 \times 10^{-4}$ (Vid. 1a);
\item Decreasing angular velocity: $v_0 = 0.02$, $n = 1/2$, giving $\Omega(\Delta r)=9.2 \times 10^{-4} r^{-1/2}$ (Blue line in Fig. 1c and Vid. 1b);
\item Increasing angular velocity: $v_0 = 0.1, n = 2$, giving $\Omega(\Delta r)=3.6 \times 10^{-6} r$ (Orange line in Fig. 1c and Vid. 1c);
\item Increasing angular velocity: $v_0 = 0.2, n = 2$, giving $\Omega(\Delta r)=7.2 \times 10^{-6} r$ (Figs. 1a-1b, Green line in Fig. 1c and Vid. 1d).
\end{itemize}

\subsection{Structured vortices in Fig. 2 and Fig. \ref{figS:PhasPort}}

The structured flows in Fig. 2 contain regions with active vortices and regions with nearly zero velocities. To generate $\textbf{v}^{tar}$ for these structured flows, we divide the system into four equal-sized squares and assign $\textbf{v}^{tar}$ values to each square independently.

For example, to generate the flow field in Fig. 2a in the main text, we set nonzero $\textbf{v}^{tar}$ values to the lower-left square, while $\textbf{v}^{tar}$ is zero in the other three squares. The lower-left square, containing a clover-shaped vortex, is assigned a velocity field: $v^{tar}_{x} = - v_1 \Delta y/\Delta r - k v_1 \sin(n \theta - \theta_c) \Delta x / \Delta r$ and $v^{tar}_{y} =  v_1 \Delta x/\Delta r - k v_1 \sin(n \theta - \theta_c) \Delta y/\Delta r$, where $v_1=v_0 \sin(\pi \Delta r / r_0)$. The first terms in $v^{tar}_{x}$ and $v^{tar}_{y}$ represent circular flow with rotational symmetry, while the second terms introduce anisotropy to the flow field with $k=2$ a coefficient used to adjust its relative weight. $n=4$ is the number of leaves, $\theta=\text{atan2}(\Delta y, \Delta x)$ and the rotation angle $\theta_c=0$. Similar to the circular flow, $\Delta r = |\textbf{r} - \textbf{r}_c|$, $\Delta x = (\textbf{r} - \textbf{r}_c)_x$, $\Delta y = (\textbf{r} - \textbf{r}_c)_y$. We set the vortex center $\textbf{r}_c=[L_x/4, L_y/4]$, vortex size $r_0=L_x/4$, and flow-strength parameter $v_0=1$. 

Such a flow field $\textbf{v}^{tar}$, after enforced incompressibility, gives a relatively strong flow in the other three squares, which are supposed to have zero flow. So, the following iterative procedure is used to solve this problem:

\begin{itemize}
    \item Step 1: Construct a preliminary target velocity field \(\textbf{v}^{tar}\) as described above.
    \item Step 2: Enforce the condition of incompressibility by using the vortex-stream function method, which gives a new flow field.
    \item Step 3: Manually set the velocities in the other squares to be zero.
    \item Step 4: Same as step 2.
    \item Repeat steps 3 and 4 until the velocities in the other squares are nearly zero.
\end{itemize}

Similar to Figs. 2a-2c, there is a single vortex in Fig. 2g. So, the same procedure is used here, except that the clover-shaped vortex is replaced by a square vortex. The latter has the following flow field: $v^{tar}_{x} = -v_1 \Delta y/\Delta r + v_2 \Delta x/\Delta r \cdot \text{sign} (\text{mod}(n \theta, 2 \pi) - \pi) \cdot \cos{(n \theta)}$ and $v^{tar}_{y} = v_1 \cdot \Delta x/\Delta r + v_2 \cdot \Delta y/\Delta r \cdot \text{sign} (\text{mod}(n \theta, 2\pi) - \pi) \cdot \cos{(n \theta)}$. Here, $v_1=v_0 \cdot \sin{\pi \Delta r/r_0}$, $v_2=0.5 v_1 \cdot (\text{sign}(\Delta r-0.5 r_0)+1) \cdot \sin{((\text{sign}(\Delta r-0.5 r_0)+1)/2)}$ are the magnitudes of the isotropic and anisotropic components of the flow, respectively. $r_0$ is set to be $L_x/2$, and the flow-strength parameters are: $v_0=1, 3, 5$ for the blue, red, orange lines.

In Figs. 2d-2f, 2h-2i, and \ref{figS:PhasPort}, the system contains two square vortices. So, we prescribe square vortices to the lower-left and top-right squares, while leaving the rest of the system flowless. The two square vortices only differ in the flow-strength parameter $v_0$. Once the initial flow field is assigned, we execute the iterative procedure mentioned before to enforce incompressibility and nearly zero flow in the top-left and lower-right squares.

\subsection{Structured vortices in Fig. 3, Fig. \ref{figS:lattice}, and Fig. \ref{figS:disorder}}

The flow fields in Figs. 3a-3c are similar to the circular vortices in Fig. 1 (see Sec. S7.A), but with different structure parameters. Specifically, we have $L_x=L_y=512$, and $v_0=1$ for Figs. 3a-3b. 

The procedure to generate vortex lattices depends on the specific lattice structure. To obtain the vortex lattice in Fig. 3d and Figs. S5g-5i, we divide the system into a \(4\times4\) grid of squares. Then, in each square, we initiate a clover-shaped vortex (see Sec. S7.B), before enforcing incompressibility. The vortices in Fig. 3d and Figs. S5g-5i are identical, with the following parameters: $L_x=L_y=256$, $r_0=L_x/8$, $\theta_c=0$, $n=4$, $k=2$, and $v_0=1$.

The vortex lattices in Figs. S5a-5f are generated using a simpler method. Instead of generating the velocity field, we prescribe a periodic vorticity field $\omega(\textbf{r})$, and then use the vortex-stream function method to obtain the corresponding velocity field $\textbf{v}(\textbf{r})$. The vorticity field we used has the following form $\omega(x,y) = \omega_0 \cdot \sin{(2\pi k_xx/L_x)} \cdot \sin{(2\pi k_yy/L_y)}$, where $\omega_0$ represents the characteristic vorticity. $k_x$ and $k_y$ indicate the number of periods along the x and y directions, respectively. The specific parameters are $\omega_0=1, k_x=1, k_y=2$ for simulations in Figs. S5a-5c; and $\omega_0=1, k_x=2, k_y=2$ for simulations Figs. S5d-5e.

Generation of disordered flows in Figs. 3e-3f and \ref{figS:disorder} follows very similar procedure. We first generate a random vorticity field $\omega_1(x, y)$ where the value on each grid is randomly picked from a Normal distribution with zero mean and unit variance, \(N(0,1)\). This gives a field of random white noise without spatial structure. To create structures, we apply a wavenumber filter in Fourier space. Specifically, we Fourier transform $\omega_1(x, y)$ to get its Fourier amplitudes $\hat{\omega}_1(k_x, k_y)$, and filtering out the long-wave and short-wave components using $\hat{\omega}(k_x, k_y)=\hat{\omega}_1(k_x, k_y)\cdot F(|k|)$. $|k|=\sqrt{k_x^2+k_y^2}$ is the wavenumber and $F(|k|)=\omega_0 k_0^{4} \cdot k^{10}/(k_0^{14} + k^{14})$ is the filter with a characteristic wavenumber $k_0$ and a typical vorticity $\omega_0$. We then perform an inverse Fourier transform to get the vorticity field in real space $\omega(x,y)$. The resulting random flow used in Figs.~3e--3f has a maximum velocity of approximately $1.4$, comparable to the characteristic velocity $v_0=1$ of the vortex array in Fig.~3d.

The flow fields in Figs. 3g-3i are prescribed using the same method as the circular vortices, except that the initial flow structure takes the following form: $v^{tar}_{x} = -v_1 \Delta y/\Delta r + v_2 \Delta x/\Delta r \cdot \text{sign} (\text{mod}(n \theta, 2 \pi) - \pi) \cdot \cos{(n \theta)}$ and $v^{tar}_{y} = v_1 \cdot \Delta x/\Delta r + v_2 \cdot \Delta y/\Delta r \cdot \text{sign} (\text{mod}(n \theta, 2\pi) - \pi) \cdot \cos{(n \theta)}$. Again, $v_1=v_0 \cdot \sin{\pi \Delta r/r_0}$ and $v_2=0.5 v_0 \cdot (\text{sign}(\Delta r-0.5 r_0)+1) \cdot \sin{((\text{sign}(\Delta r-0.5 r_0)+1)/2)}$ are the magnitudes of the isotropic and anisotropic components of the flow, respectively. We set $n=2, r_0=L_x/2$. The flow speeds are $v_0=0.01$ for Fig. 3g, $v_0=10$ for Fig. 3h, and in Fig. 3i, $v_0$ varies from 0 to 10.

\subsection{Flows in Fig. \ref{figS:align}}

The flows in Figs. \ref{figS:align}a-\ref{figS:align}c have the same structure as the clover-shaped vortex in Fig. 2a, but differ in the flow-strength parameter \(v_0\). Specifically, we have $v_0=0.001$ for Fig. S3a, $v_0=0.004$ for Fig. S3b and $v_0=0.1$ for Fig. S3c. 

The elliptical vortex in Fig. \ref{figS:align}d is generated using $v^{tar}_{x} = - v_0 \Delta y/b$ and $v^{tar}_{y} = v_0 \Delta x/a$, where $a$ and $b$ represent the major and minor axes of the ellipse, respectively. The parameters are $v_0=3, a=L_x/2, b=L_y/3$.

We generate the capsule-shaped vortex in Fig. \ref{figS:align}e by combining a rectangular flow field with two semicircular flow fields. To do this, we define two parameters: the distance between the centers of the two semicircles $l_0$, and the radius of the semicircles $r_0$. The rectangular flow resides in the rectangular box $L_x/2-l_0/2\le x\le L_x/2+l_0/2$ and $L_y/2-r_0\le y\le L_y/2+r_0$. In this region, we set $v_x=-v_0 \cdot \sin{\pi \Delta y/r_0}, v_y=0$. Right next to the rectangular region, we have two semicircular flows attached to the left and right sides, respectively. The centers of the two semicircles are $[L_x/2-l_0/2, L_y/2]$ and $[L_x/2+l_0/2, L_y/2]$. In the semicircle, the velocity field is set as $v_x = -v \cdot \Delta y/r_0, v_y = v \cdot \Delta x/r_0$, where  $v=v_0 \cdot \sin{\pi \Delta r/r_0}$. The parameters used in Fig. S3e are $v_0=3$ and $r_0=L_x/4$.

Finally, the six-gear star in Fig. \ref{figS:align}f has the same expression of velocity as the vortices in Figs. \ref{figS:align}a-\ref{figS:align}c, but the structure parameters are slightly different: $n=6, k=2$ and $v_0 = 3$. 



\section{Robustness tests}
\label{secS:robustness}

We examined the robustness of the two-droplet coexistence states against
stochastic perturbations, variations in initial droplet positions,
numerical refinement, mobility contrast, and weak flow asymmetry. Unless
otherwise stated, both equal-size and unequal-size coexistence were
tested. Imperfect and time-dependent velocity fields are considered
separately using the coupled CHNS model in Sec.~S2.

\begin{figure}[t]
    \centering
    \includegraphics[width=0.7\linewidth]{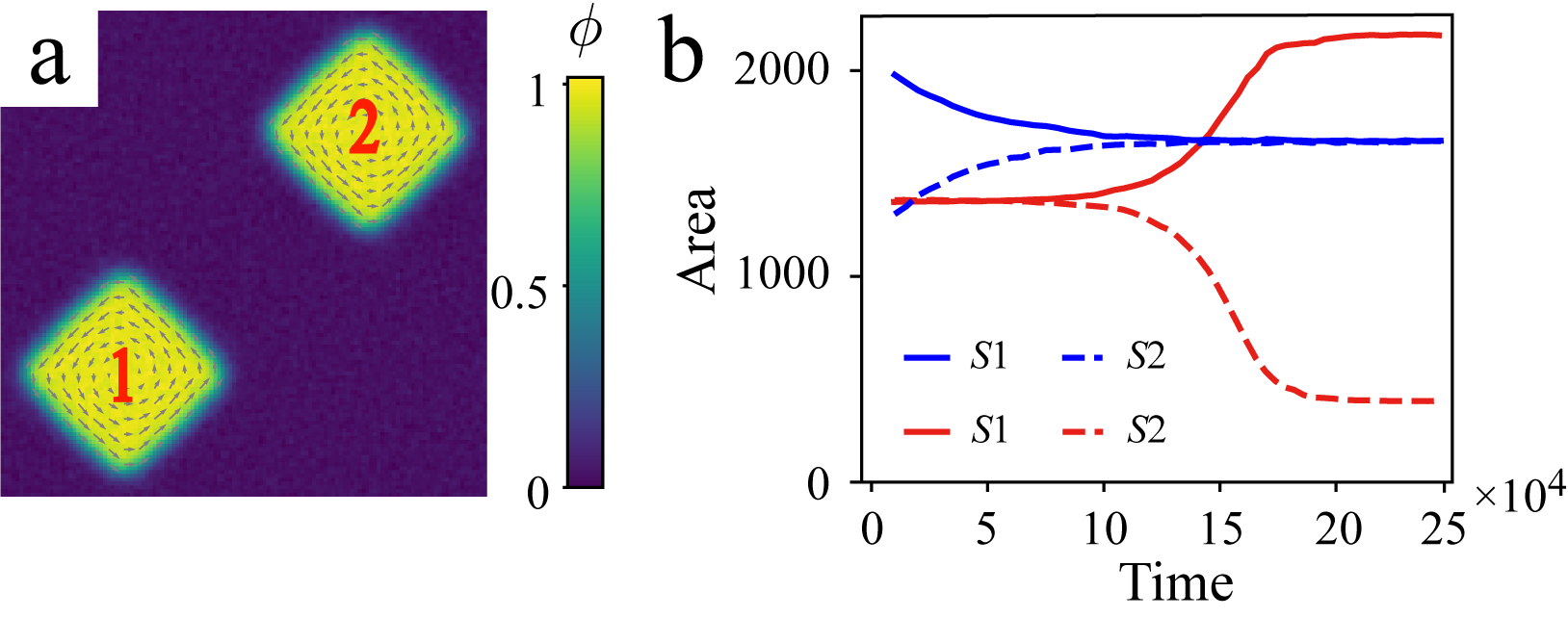}
    \caption{\textbf{Robustness against stochastic perturbations.}
    (a) Representative configuration under a chemical-potential perturbation with amplitude $A_\mu=0.1$. (b) Evolution of $S_1$ (solid) and $S_2$ (dashed) for
    equal-size (blue) and unequal-size (red) coexistence.}
    \label{figS:noise}
\end{figure}

First, we tested robustness against a spatially uncorrelated Gaussian perturbation of the chemical potential,\begin{equation}\mu_i^n=\mu_{\mathrm{det},i}^n+A_\mu\xi_i^n,\end{equation}where $\xi_i^n$ is independently sampled from a standard normal distribution at each grid point and time step, and $A_\mu=0.1$. Because the perturbation enters through $M\nabla^2\mu$, the total composition remains conserved under periodic boundary conditions. Figure~\ref{figS:noise} shows that both coexistence states persist under the tested noise realization, with only small fluctuations in the droplet areas.

\begin{figure}[t]
    \centering
    \includegraphics[width=0.7\linewidth]{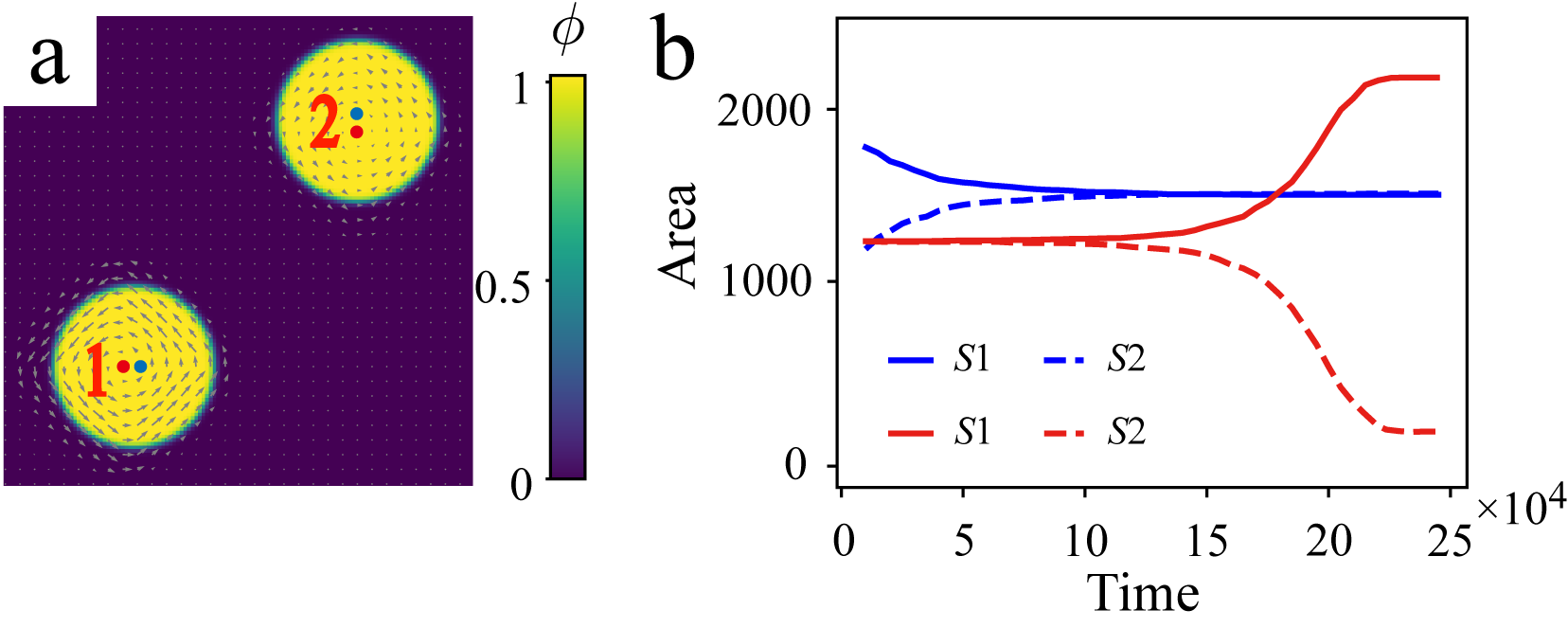}
    \caption{\textbf{Robustness against variations in initial droplet positions.} (a) Snapshot of the initial configuration at $t=0$, in which each droplet is displaced from its corresponding flow center by $10\%$ of the vortex radius. Red dots mark the centers of the prescribed flows, and blue dots indicate the initial droplet centers of mass. (b) Evolution of $S_1$ (solid) and $S_2$ (dashed)  for equal-size (blue) and unequal-size (red) coexistence.}
    \label{figS:position}
\end{figure}

We next displaced each initial droplet independently from its vortex
center by $10\%$ of the vortex radius. The droplets are recaptured by
the corresponding flows and recover the equal-size or unequal-size
coexistence states (Fig.~\ref{figS:position}), showing that precise
initial positioning is not required.

\begin{figure}[t]
    \centering
    \includegraphics[width=0.7\linewidth]{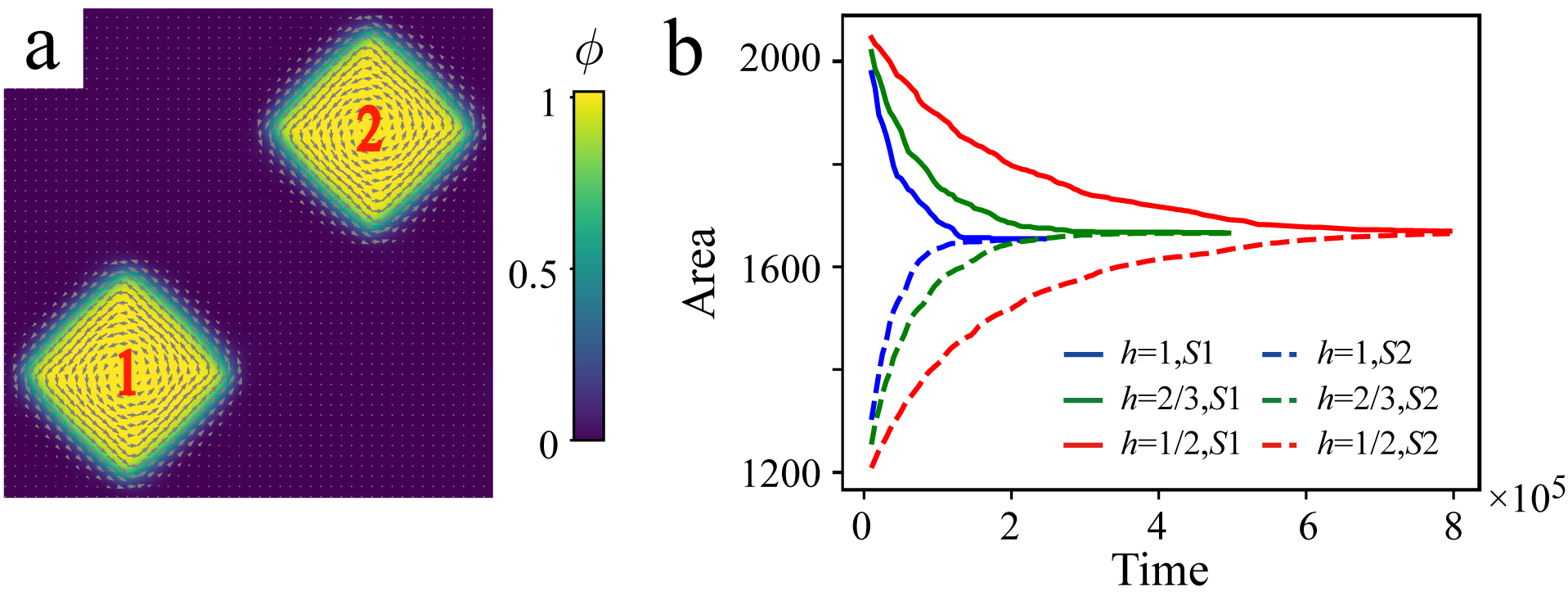}
    \caption{\textbf{Numerical-resolution test corresponding to
    Fig.~2d.}
    (a) Late-time configuration on the $192^2$ grid.
    (b) Evolution of $S_1$ (solid) and $S_2$ (dashed) on $128^2$
    (blue), $192^2$ (green), and $256^2$ (red) grids.}
    \label{figS:grid}
\end{figure}

For the numerical-resolution test, we fixed the physical domain size at
$L_x=L_y=128$ and used
$(h,\Delta t)=(1,0.1)$, $(2/3,0.05)$, and $(1/2,0.001)$, corresponding
to $128^2$, $192^2$, and $256^2$ grids, respectively. Although the
transient relaxation rates differ, all three simulations show the same
direction of material transfer and tendency toward size equalization
(Fig.~\ref{figS:grid}).

\begin{figure}[t]
    \centering
    \includegraphics[width=0.7\linewidth]{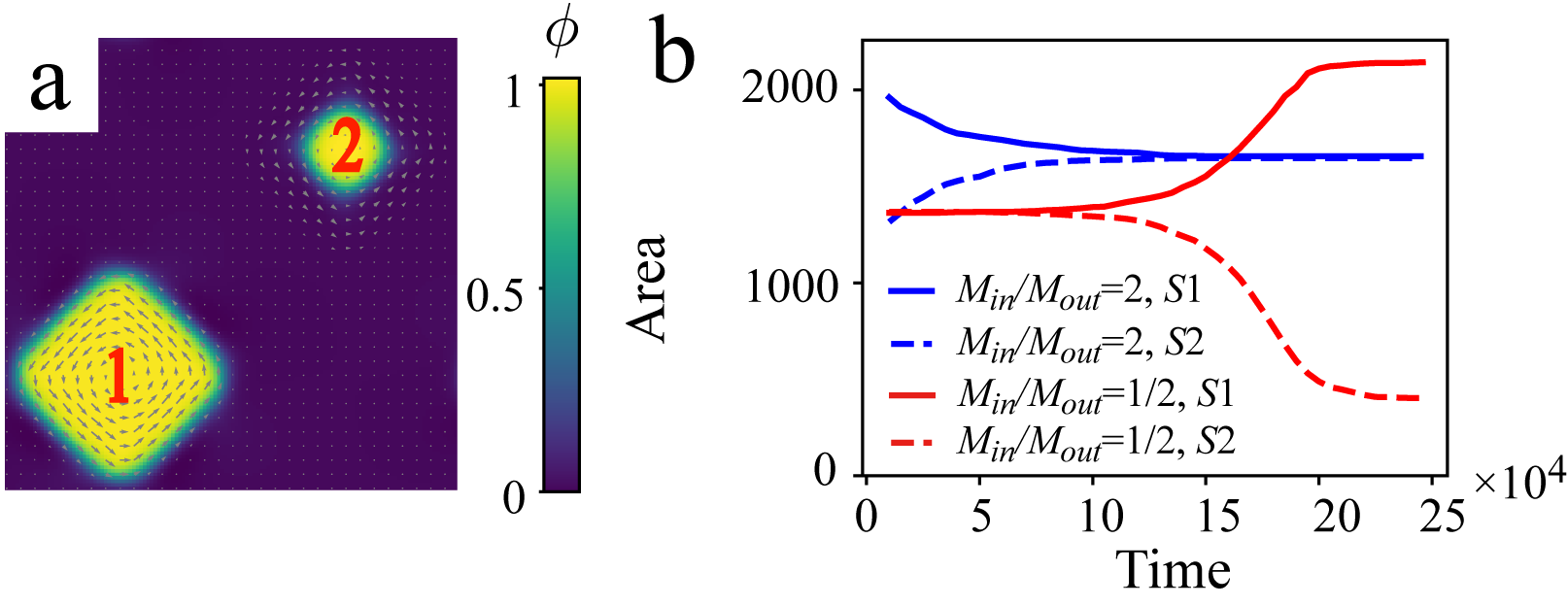}
    \caption{\textbf{Robustness against mobility contrast.}
    (a) Representative late-time configuration.
    (b) Evolution of $S_1$ (solid) and $S_2$ (dashed) for
    $M_{in}/M_{out}=2$ and $0.5$.}
    \label{figS:mobility}
\end{figure}

To examine the mobility contrast, we used a composition-dependent
mobility,
\begin{equation}
    M(\phi)=M_{out}(1-\phi)+M_{in}\phi,
\end{equation}
with $M_{in}/M_{out}=2$ and $0.5$. Both equal-size and unequal-size
coexistence remain stable, while the relaxation rates change
(Fig.~\ref{figS:mobility}). Since viscosity does not enter the
prescribed-flow model, viscosity-dependent effects are instead examined
through the CHNS simulations in Sec.~S2.

\begin{figure}[t]
    \centering
    \includegraphics[width=0.7\linewidth]{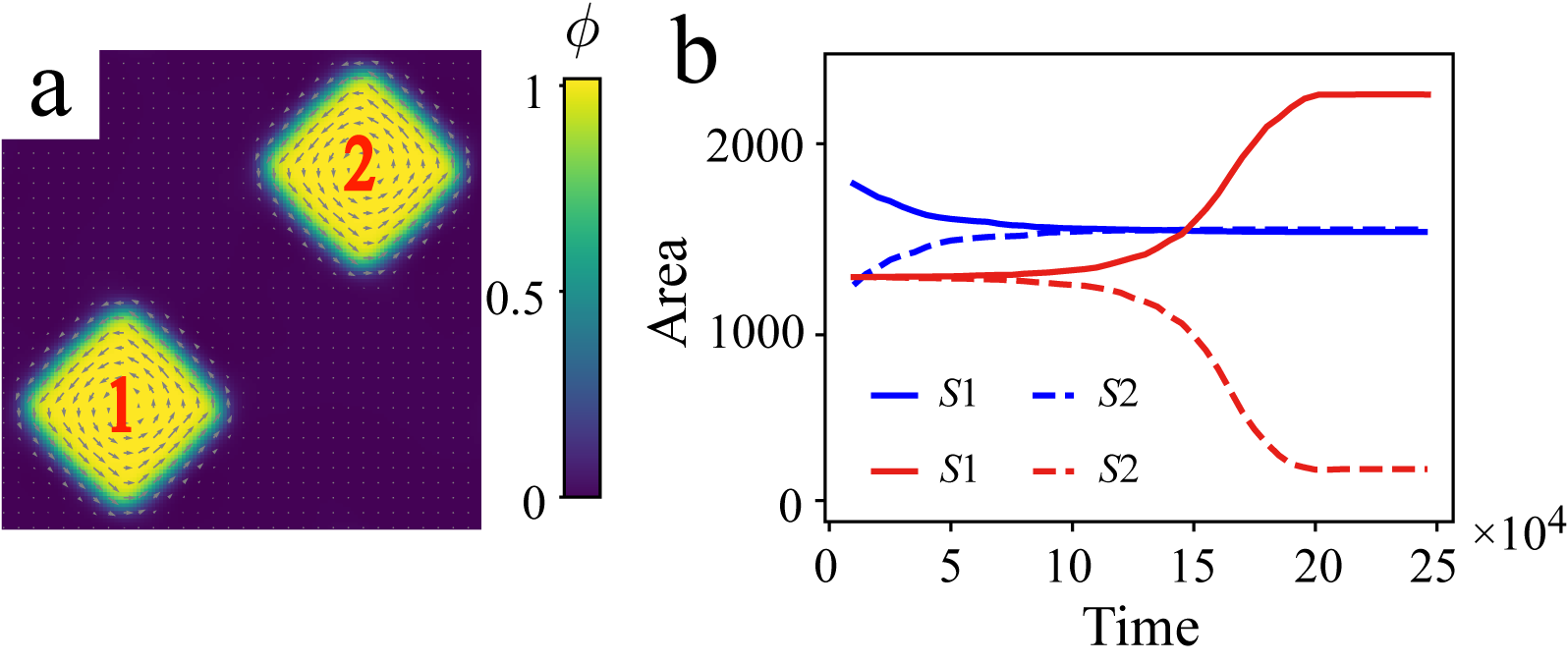}
    \caption{\textbf{Robustness against weak flow asymmetry.}
    (a) Representative configuration under a weak flow asymmetry with \(\epsilon=0.05\). (b) Evolution of $S_1$ (solid) and $S_2$ (dashed)
    for equal-size (blue) and unequal-size (red) coexistence.}
    \label{figS:asymmetry}
\end{figure}

Finally, we introduced a weak flow asymmetry by multiplying the velocity amplitude of each vortex by $1+\varepsilon\cos(\theta-\theta_a)$, with $\varepsilon=0.05$ and different asymmetry directions $\theta_a$ for the two vortices. This strengthens one side of each vortex by up to $5\%$ and weakens the opposite side by the same amount, while preserving the angularly averaged flow amplitude. The resulting field was projected to ensure incompressibility. The equal-size and unequal-size coexistence behaviors remain qualitatively unchanged
(Fig.~\ref{figS:asymmetry}).

Together, these tests show that small physical and numerical perturbations may alter the relaxation kinetics but do not change the flow-selected coexistence or the qualitative direction of material transfer.

\newpage
\section{Videos}

\textbf{Vid. 1}: A single droplet in circular vortices with different angular velocities: (a) $\Omega(r)=4.2 \times 10^{-4}$, (b) $\Omega(r)=9.2 \times 10^{-4} r^{-1/2}$, (c) $\Omega(r)=3.6 \times 10^{-6} r$, and (d) $\Omega(r)=7.2 \times 10^{-6} r$.

\textbf{Vid. 2}: Two droplets each driven by a patterned flow. (a) No flow in the background, corresponding to thermal equilibrium. (b-d) Corresponding to Figs. 2a, 2d, 2e in the main text, respectively.

\textbf{Vid. 3}: Dynamic manipulation of the droplet sizes. Panels a and b correspond to the blue and red lines in Fig. 2i in the main text.

\textbf{Vid. 4}: Phase separation (a) without and (b) with a structured flow in the background. Panel b corresponds to Fig. 3a in the main text.

\textbf{Vid. 5}: Phase separation driven by a periodic array of vortices, corresponding to Fig. 3d in the main text.

\textbf{Vid. 6}: Phase separation driven by a static disordered flow, corresponding to Fig. 3e in the main text.

\textbf{Vid. 7}: Melting of droplets by structured vortices. Panels a and b correspond to Figs. 3g and 3h in the main text. 

\textbf{Vid. 8}: Self-sustained oscillations of (a) a single droplet in a clover-shaped flow field with $v_0$ = 0.04, and (b) a lattice of droplets in a vortex array with initial $\phi_0=0.3$ and $\omega_0=0.1$.

\bibliographystyle{apsrev4-1}
%